\documentclass[useAMS,usenatbib]{mn2e}
\usepackage{graphicx}
\title[A first orbit for R145]{A first orbital solution for the very
massive 30 Dor main-sequence WN6h+O binary R145} \author[O. Schnurr
et al.]{O. Schnurr$^{1,2}$\thanks{Visiting Astronomer at the Complejo
Astron\'omico El Leoncito (CASLEO), the Cerro Tololo Inter-American
Observatory (CTIO), the South African Astronomical Observatory (SAAO),
and the Mount Stromlo and Siding Spring Observatory
(MSSSO)}\thanks{E-mail: o.schnurr@sheffield.ac.uk},
A. F. J. Moffat$^{1}$, A. Villar-Sbaffi$^{1}$, N. St-Louis$^{1}$,
\newauthor and N. I. Morrell$^{3}$ \vspace{3mm}\\ $^{1}$Dept. de
Physique, Universit\'e de Montr\'eal, C. P. 6128, succ. centre-ville,
Montr\'eal (Qc) H3C 3J7, and Centre de Recherche\\en Astrophysique du
Qu\'ebec, Canada\\ $^{2}$Dept. of Physics and Astronomy, University of
Sheffield, Hicks Building Hounsfield Road, Sheffield S3 7RH, United
Kingdom\\ $^{3}$Las Campanas Observatory, Observatories of the
Carnegie Institution of Washington, Casilla 601, La Serena, Chile}

\begin{document}

\date{Version 18 December 2008}

\pagerange{\pageref{firstpage}--\pageref{lastpage}} \pubyear{2008}

\maketitle

\label{firstpage}

\begin{abstract}

We report the results of a spectroscopic and polarimetric study of the
massive, hydrogen-rich WN6h stars R144 (HD 38282 = BAT99-118 = Brey
89) and R145 (HDE 269928 = BAT99-119 = Brey 90) in the LMC. Both stars
have been suspected to be binaries by previous studies (R144:
\citealt{S08b}; R145: \citealt{M89}). We have combined radial-velocity
(RV) data from these two studies with previously unpublished
polarimetric data. For R145, we were able to establish, for the first
time, an orbital period of 158.8 days, along with the full set of
orbital parameters, including the inclination angle $i$, which was
found to be $i = 38^{\circ} \pm 9^{\circ}$. By applying a modified
version of the shift-and-add method developed by \citet{Demers02}, we
were able to isolate the spectral signature of the very faint-line
companion star. With the RV amplitudes of both components in R145, we
were thus able to estimate their absolute masses. We find minimum
masses M$_{\rm WR}sin^{3}i = (116 \pm 33)$ M$_{\odot}$ and $M_{\rm
O}sin^{3}i = (48 \pm 20)$ M$_{\odot}$ for the WR and the O component,
respectively. Thus, if the low inclination angle were correct,
resulting absolute masses of the components would be at least 300 and
125 M$_{\odot}$, respectively. However, such high masses are not
supported by brightness considerations when R145 is compared to
systems with known, very high masses such as NGC3603-A1 or WR20a. An
inclination angle close to 90$^{\circ}$ would remedy the situation,
but is excluded by the currently available data. More and better data
are thus required to firmly establish the nature of this puzzling, yet
potentially very massive and important system. As to R144, however,
the combined data sets are not sufficient to find any periodicity.

\end{abstract}


\section{Introduction}


Fundamental questions of stellar astrophysics suffer from a lack of
truly empirical evidence when it comes to stars with the highest
masses both on and evolved off the main sequence (MS). Recent models
maintain that in the early Universe, the initial-mass function (IMF)
was top-heavy and that primordial (Population III) stars were
extremely massive, with masses from $\sim$100 to $\sim$1000 M$_{\sun}$
(\citealt{Ostriker96}; \citealt{Nakamura01}; \citealt{Schaerer02}). In
contrast to that, recent studies have put forward both theoretical and
observational arguments that under present-day conditions, an IMF
cut-off occurs around $\sim150$ M$_{\sun}$ (\citealt{WeidKroup04};
\citealt{Figer05}). However, even in the Local Group very massive
stars remain poorly characterized, and it so far remains uncertain at
what mass an upper cut-off really occurs, if at all.


The least model-dependent and thus most reliable way to directly
measure stellar masses is by Keplerian orbits in binary systems. In
the past years, studies of very massive binaries have yielded the
somewhat surprising results that the most massive stars known are to
be found not among O-type stars, but among a very luminous subtype of
Wolf-Rayet (WR) stars, the so-called WN5-7h (or ha) stars. These
hydrogen-rich WN stars are young, unevolved objects rather than
``classical'' WR stars, which are usually identified with evolved,
helium-burning, massive stars (\citealt{deKoter97};
\citealt{CroDess98}). WN5-7h stars mimic classical, i.e. evolved, core
helium burning WR stars because their extreme luminosity, a result of
their very high masses, drives a dense and fast wind which gives rise
to WR-like emission lines. Due to the rarity of very massive stars in
general and of very massive binaries in particular, only few massive
systems have been ``weighed'' so far. In order of increasing mass,
these are the Galactic Wolf-Rayet stars WR22 (72 M$_{\sun}$ for the
WN7h component \citep{Rauw96b}, although \citet{Schweick99} derive
only 55 M$_{\sun}$), WR20a (83 and 82 M$_{\sun}$ for both WN6ha
components; \citealt{Rauw04}; \citealt{Bonanos04}), WR21a (minimum
mass $M\sin^{3}i = 87$ M$_{\sun}$ for the O3If/WN6 primary;
\citep{Niemela08}), NGC3603-A1 (116 and 89 M$_{\sun}$ for both WN6ha
components; \citealt{S08a}). Clearly, these very high masses
impressively underline the extraordinary nature of WN5-7h stars.

In a recent, intense survey of 41 of the 47 known WNL stars in the
LMC, \citet[hereafter S08]{S08b} used 2m-class telescopes to obtain
repeated, intermediate-quality (R=1000, S/N$\sim$80) spectra in order
to assess the binary status of the 41 targets from radial-velocity
(RV) variations. S08 identified four new binaries containing
WN5-7h stars, bringing the total number of known WNLh binaries in the
LMC to nine. For one of the previously known binaries, R145 (= HDE
269928 = BAT99-119 = Brey 90), the preliminary 25.4-day period
reported by \citet[hereafter M89]{M89} could not be confirmed, but
neither could a coherent period be established from the RV data of
this study alone. Therefore, we combined S08's RVs with those
published by M89, and with previously unpublished polarimetry. It was
only after the combination that we were able to find the true orbital
period of this system as well as other orbital parameters.

From both its RV and EW variations and its high X-ray luminosity, S08
identified another WN6h star, R144 (HD 38282 = BAT99-118 = Brey 89),
to be a binary candidate, although M89 had identified this star as
probably single. In terms of spectral type, R144 is almost a perfect
clone of R145; however, at the $\sim$same bolometric correction and
after allowing for differential interstellar extinction, R144 is
$\sim$0.5 mag brighter than R145. Indeed, \citet{CroDess98}, using
atmosphere models without iron-line blanketing, derived a
spectroscopic mass in excess of 100 M$_{\sun}$, and a luminosity of
log$L/L_{\sun} = 6.34$ for R144, which makes this star one of the most
luminous WR stars known. New, updated models most likely would yield
an even higher mass and luminosity, making R144 the most luminous
main-sequence (MS) object in the Local Group. Since we also have
unpublished polarimetry for R144, we have revisited this object as
well, using the spectroscopic data obtained by S08.

In the present paper, we will report the results of this study. In
Section \ref{section2}, we will briefly describe the observations and
the data reduction. In Section \ref{section3}, we will describe the
analysis of our data and present the results. The paper is then
summarized in Section \ref{section4}.


\section{Observations and Data Reduction}
\label{section2}

\subsection{Spectroscopy}

The spectroscopic observations are described in detail in S08. We
briefly recapitulate here that long-slit spectrographs attached to
ground-based, southern, 2m-class telescopes were used. Data were
obtained in three observing seasons between 2001 and 2003 to
maximize the time coverage, and were carried out during 13 runs at 6
different telescopes. The spectral coverage varied from one instrument
to another, but all spectrographs covered the region from 4000 to 5000
\AA, thereby encompassing the strategic emission line He\,\textsc{ii}
$\lambda$4686

Data were reduced in the standard manner using NOAO-IRAF\footnote{IRAF
is distributed by the National Optical Astronomy Observatories, which
are operated by the Association of Universities for Research in
Astronomy, Inc., under cooperative agreement with the National Science
Foundation.}, and corrected for systematic shifts among different
observatories (see S08 for more details). The achieved linear
dispersion varied from 0.65 \AA/pixel to 1.64 \AA/pixel, but all data
were uniformly rebinned to 1.65 \AA/pixel, thereby yielding a
conservative 3-pixel resolving power of $R\sim1000$. The achieved
signal-to-noise (S/N) ratio was $\sim$ 120 per resolution element for
both stars, as measured in the continuum region between 5050 and
5350 \AA.

Additional RV data for R144 and R145 were taken from M89, to which we
refer for more details on the data acquisition and reduction.

\subsection{Polarimetry}

Linear polarimetry in white light was obtained during a total of six
observing runs between October 1988 and May 1990 at the 2.2m telescope
of the Complejo Astron\'omico El Leoncito (CASLEO) near San Juan,
Argentina, with VATPOL (\citealt{Magalhaes84}), and at the ESO/MPG-2.2m
telescope at La Silla, Chile, with PISCO (\citealt{Stahl86}). RCA GaAs
phototubes were employed in both cases. Exposure times were typically
15 minutes per data point. Appropriate standard stars were taken to
calibrate the polarization angle and the zero level, and exposures of
the adjacent sky were obtained to correct for the background count
rates.

The polarimetric data were calibrated the usual way
(e.g. \citealt{Villar06}). Since Thompson scattering is
wavelength-independent (i.e. grey), and the differences between the
sensitivity curves of the respective detectors are small, no
correction for the different passbands was deemed necessary. However,
a small instrumental offset between PISCO and VATPOL was found in
polarization; we therefore have added $\Delta Q = 0.10\%$ and $\Delta
U = 0.10\%$ to the PISCO data to match the VATPOL data, on
average. The final polarimetric data for R144 and R145 are listed in
Tables \ref{tablepol118} and \ref{tablepol119}, respectively.

\begin{figure}
\includegraphics[width=63mm,angle=-90,trim= 0 30 0 20,clip]{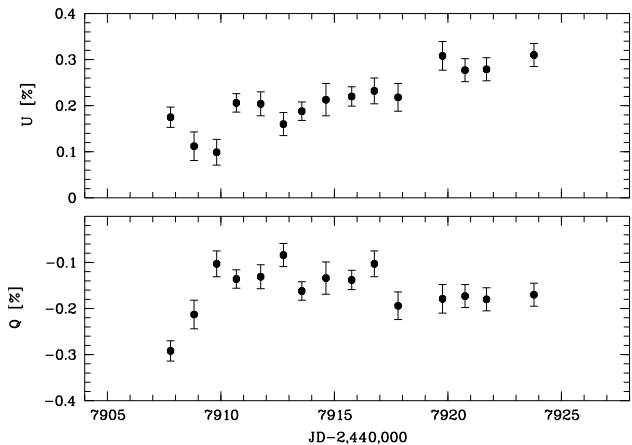}
\caption{Polarimetric data for R144 plotted versus Julian date.}
\label{poldata118}
\end{figure}

Statistics reveal that R144 displays a small but significant scatter
(standard deviation) about their mean values in both Stokes
parameters, 0.052\% and 0.063\% in $Q$ and $U$, respectively; by
comparison, the quoted error per data point is only $\sim$half
this. However, in Figure \ref{poldata118} one can clearly see that
over the observation period of $\sim$15 days, the polarization in
Stokes-$U$ increases constantly by $\sim$0.2\% (i.e. $\sim$seven times
the quoted error per data point), while Stokes-$Q$ also shows
systematic variability, which is a strong indication that some
coherent process is involved, e.g. a (longer-period) binary motion.

\begin{figure}
\includegraphics[width=63mm,angle=-90,trim= 0 30 0 20,clip]{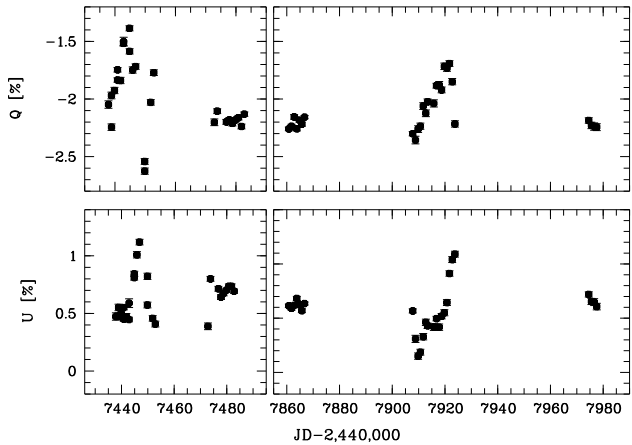}
\caption{Polarimetric data for R145 plotted versus Julian date. Note the
enormous amplitude of $\sim$1\%, rarely seen so large in other WR+O
systems.}
\label{poldata119}
\end{figure}

R145, on the other hand, is much more variable; the scatter is 0.26\%
and 0.20\% about their mean values, respectively, in $Q$ and $U$,
with well defined and coherent variability episodes (Figure
\ref{poldata119}).

\begin{table}
\caption{Linear polarimetry for R144, obtained with VATPOL at
CASLEO over a two-week period in January 1990.}
\label{tablepol118}
\begin{tabular}{cccccc}
\hline
HJD & $Q$ & $\sigma_{Q}$ &  $U$ & $\sigma_{U}$  & Instrument\\
\hline
2447907.783   &	-0.292  &0.022 & 0.175 &0.022 & VATPOL \\
2447908.819   &	-0.213  &0.031 & 0.112 &0.031 & VATPOL \\
2447909.813   &	-0.103  &0.028 & 0.099 &0.028 & VATPOL \\
2447910.681   &	-0.136  &0.020 & 0.206 &0.020 & VATPOL \\
2447911.754   &	-0.131  &0.026 & 0.204 &0.026 & VATPOL \\
2447912.754   &	-0.084  &0.025 & 0.160 &0.025 & VATPOL \\
2447913.566   &	-0.162  &0.020 & 0.188 &0.020 & VATPOL \\
2447914.628   &	-0.134  &0.035 & 0.213 &0.035 & VATPOL \\
2447915.762   &	-0.138  &0.021 & 0.220 &0.021 & VATPOL \\
2447916.758   &	-0.103  &0.028 & 0.232 &0.028 & VATPOL \\
2447917.803   &	-0.194  &0.030 & 0.218 &0.030 & VATPOL \\
2447919.763   &	-0.179  &0.031 & 0.308 &0.031 & VATPOL \\
2447920.753   &	-0.173  &0.025 & 0.277 &0.025 & VATPOL \\
2447921.699   &	-0.180  &0.025 & 0.279 &0.025 & VATPOL \\
2447923.789   &	-0.170  &0.025 & 0.310 &0.025 & VATPOL \\
\hline
\end{tabular}
\end{table}

\begin{table}
\caption{Linear polarimetry for R145, obtained with PISCO at ESO
and VATPOL at CASLEO, between October 1988 and May 1990.}
\label{tablepol119}
\begin{tabular}{cccccc}
\hline
HJD & $Q$ & $\sigma_{Q}$ & $U$ & $\sigma_{U}$ & Instrument\\
\hline
2447438.872  &  -2.244  & 0.028 &  0.552 &  0.028  &   PISCO\\
2447440.856  &  -1.746  & 0.020 &  0.551 &  0.020  &   PISCO \\
2447442.832  &  -1.517  & 0.020 &  0.447 &  0.020  &   PISCO \\
2447444.810  &  -1.386  & 0.022 &  0.811 &  0.022  &   PISCO\\
2447446.838  &  -1.718  & 0.024 &  1.119 &  0.024  &   PISCO \\
2447449.858  &  -2.626  & 0.027 &  0.822 &  0.027  &   PISCO \\
2447449.839  &  -2.542  & 0.025 &  0.573 &  0.025  &   PISCO \\
2447451.853  &  -2.029  & 0.024 &  0.457 &  0.024  &   PISCO \\
2447452.852  &  -1.771  & 0.025 &  0.408 &  0.025  &   PISCO \\
2447437.859  &  -2.048  & 0.032 &  0.473 &  0.032  &  VATPOL\\
2447438.849  &  -1.969  & 0.029 &  0.476 &  0.029  &  VATPOL\\
2447439.843  &  -1.926  & 0.027 &  0.502 &  0.027  &  VATPOL\\
2447440.856  &  -1.835  & 0.019 &  0.451 &  0.019  &  VATPOL \\
2447441.854  &  -1.839  & 0.027 &  0.472 &  0.027  &  VATPOL\\
2447442.873  &  -1.502  & 0.038 &  0.589 &  0.038  &  VATPOL\\
2447444.848  &  -1.586  & 0.023 &  0.843 &  0.023  &  VATPOL\\
2447445.858  &  -1.748  & 0.027 &  1.009 &  0.027  &  VATPOL\\
2447472.852  &  -2.202  & 0.029 &  0.388 &  0.029  &  VATPOL\\
2447473.810  &  -2.105  & 0.021 &  0.800 &  0.021  &  VATPOL\\
2447476.819  &  -2.198  & 0.021 &  0.714 &  0.021  &  VATPOL\\
2447477.812  &  -2.181  & 0.021 &  0.642 &  0.021  &  VATPOL\\
2447478.815  &  -2.211  & 0.023 &  0.676 &  0.023  &  VATPOL\\
2447479.803  &  -2.182  & 0.018 &  0.701 &  0.018  &  VATPOL\\
2447480.824  &  -2.162  & 0.020 &  0.736 &  0.020  &  VATPOL\\
2447481.819  &  -2.238  & 0.021 &  0.736 &  0.021  &  VATPOL\\
2447482.772  &  -2.131  & 0.019 &  0.693 &  0.019  &  VATPOL\\
2447860.775  &  -2.259  & 0.016 &  0.614 &  0.016  &  VATPOL\\
2447861.785  &  -2.236  & 0.017 &  0.591 &  0.017  &  VATPOL\\
2447862.779  &  -2.155  & 0.020 &  0.618 &  0.021  &  VATPOL\\
2447863.807  &  -2.258  & 0.019 &  0.682 &  0.019  &  VATPOL\\
2447864.793  &  -2.183  & 0.017 &  0.626 &  0.017  &  VATPOL\\
2447865.715  &  -2.218  & 0.018 &  0.569 &  0.018  &  VATPOL\\
2447866.733  &  -2.157  & 0.014 &  0.635 &  0.014  &  VATPOL\\
2447907.768  &  -2.301  & 0.019 &  0.567 &  0.019  &  VATPOL\\
2447908.808  &  -2.356  & 0.033 &  0.309 &  0.033  &  VATPOL\\
2447909.802  &  -2.261  & 0.029 &  0.150 &  0.029  &  VATPOL\\
2447910.671  &  -2.236  & 0.027 &  0.186 &  0.027  &  VATPOL\\
2447911.745  &  -2.061  & 0.029 &  0.329 &  0.029  &  VATPOL\\
2447912.742  &  -2.124  & 0.029 &  0.462 &  0.029  &  VATPOL\\
2447913.379  &  -2.024  & 0.025 &  0.430 &  0.025  &  VATPOL\\
2447915.774  &  -2.038  & 0.031 &  0.418 &  0.031  &  VATPOL\\
2447916.772  &  -1.884  & 0.021 &  0.496 &  0.021  &  VATPOL\\
2447917.792  &  -1.877  & 0.029 &  0.417 &  0.029  &  VATPOL\\
2447918.713  &  -1.921  & 0.025 &  0.518 &  0.025  &  VATPOL\\
2447919.751  &  -1.716  & 0.028 &  0.550 &  0.028  &  VATPOL\\
2447920.739  &  -1.732  & 0.027 &  0.644 &  0.027  &  VATPOL\\
2447921.690  &  -1.692  & 0.025 &  0.912 &  0.025  &  VATPOL\\
2447922.733  &  -1.849  & 0.027 &  1.040 &  0.027  &  VATPOL\\
2447923.772  &  -2.216  & 0.027 &  1.090 &  0.027  &  VATPOL\\
2447974.536  &  -2.186  & 0.024 &  0.719 &  0.024  &  VATPOL\\
2447975.531  &  -2.229  & 0.028 &  0.654 &  0.028  &  VATPOL\\
2447976.579  &  -2.239  & 0.032 &  0.648 &  0.032  &  VATPOL\\
2447977.559  &  -2.244  & 0.030 &  0.607 &  0.030  &  VATPOL\\
\hline
\end{tabular}
\end{table}

\section{Data Analysis and Results}
\label{section3}

\subsection{Spectroscopic Data}
\label{supertemplate}

We here only briefly describe the methods employed to extract radial
velocities (RVs) from the spectra (for details we refer the reader to
S08 and references therein): Cross-correlation and emission-line
fitting routines programmed using the ESO-MIDAS tasks XCORR/IMA and
FIT/IMA, respectively, with an iterative approach for
cross-correlation.



\begin{figure}
\includegraphics[width=85mm,angle=0,trim= 10 0 5 0,clip]{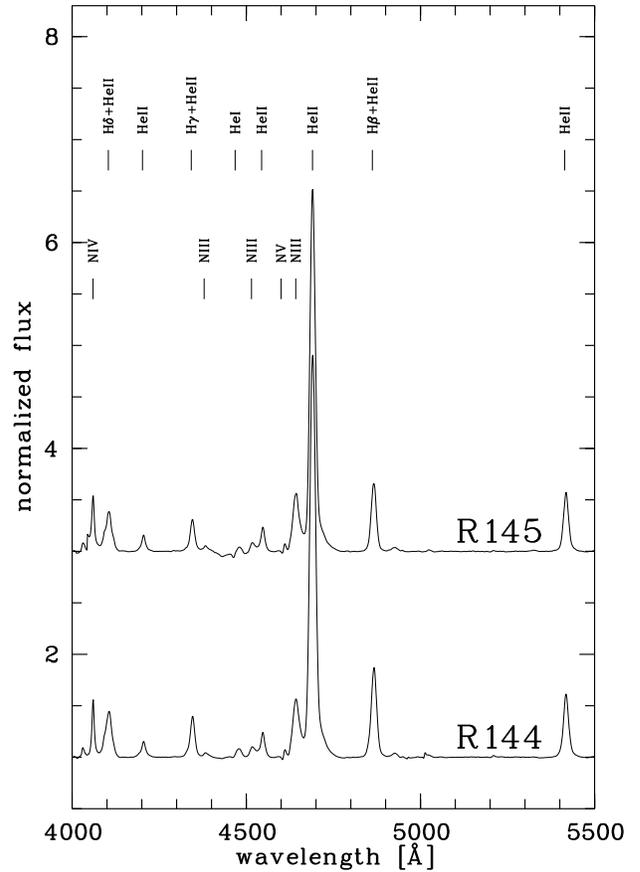}
\caption{Mean spectra of R144 and R145. For clarity, the upper
spectrum has been shifted by 2 flux units.}
\label{avespecs}
\end{figure}

Since the spectra of both R144 and R145 are largely dominated by the
strong He\,\textsc{ii} $\lambda$4686 emission line (see Figure
\ref{avespecs}), we initially confined cross-correlation to this line
to derive RVs for R145, but then applied the method to the
He\,\textsc{ii} $\lambda$5412 line as well. Furthermore, RVs from the
faint, but very narrow N\,\textsc{iv} $\lambda$4058 emission were
obtained by fitting Gauss profiles to the line.

Unfortunately, due to low S/N and an unreliable wavelength calibration
at the blue end of our spectra (the comparison-arc lamp has
practically no useful lines in this region; see S08 for more details),
the N\,\textsc{iv} $\lambda$4058 line yielded much worse RV scatter than
the two He\,\textsc{ii} lines. Thus, the main part of the analysis was
carried out with the two, much stronger He\,\textsc{ii} lines


\begin{figure*}
\begin{minipage}{170mm}
\includegraphics[width=80mm,angle=0,trim= 20 0 0 0,clip]{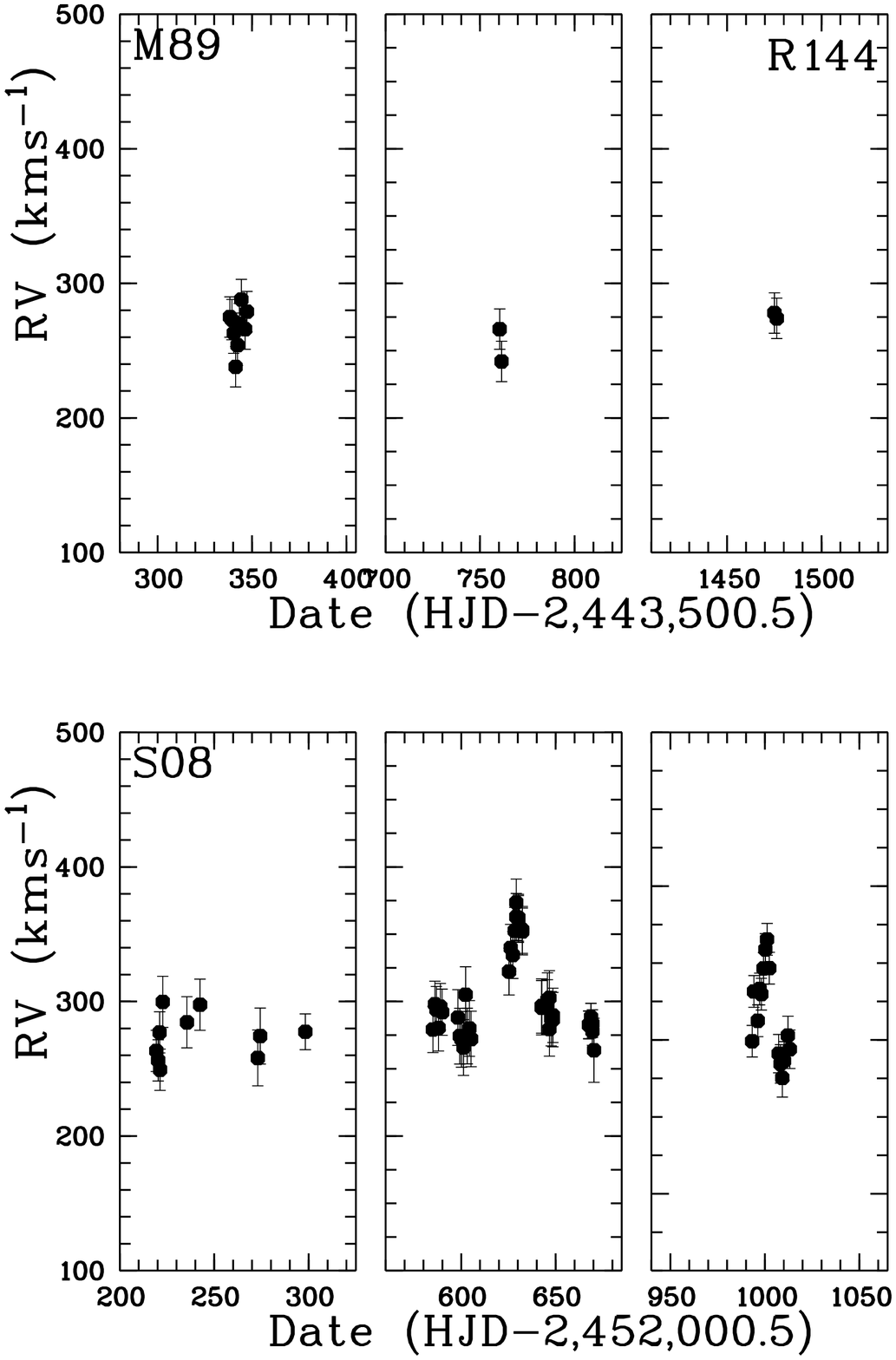}\hfill
\includegraphics[width=80mm,angle=0,trim= 20 0 0 0,clip]{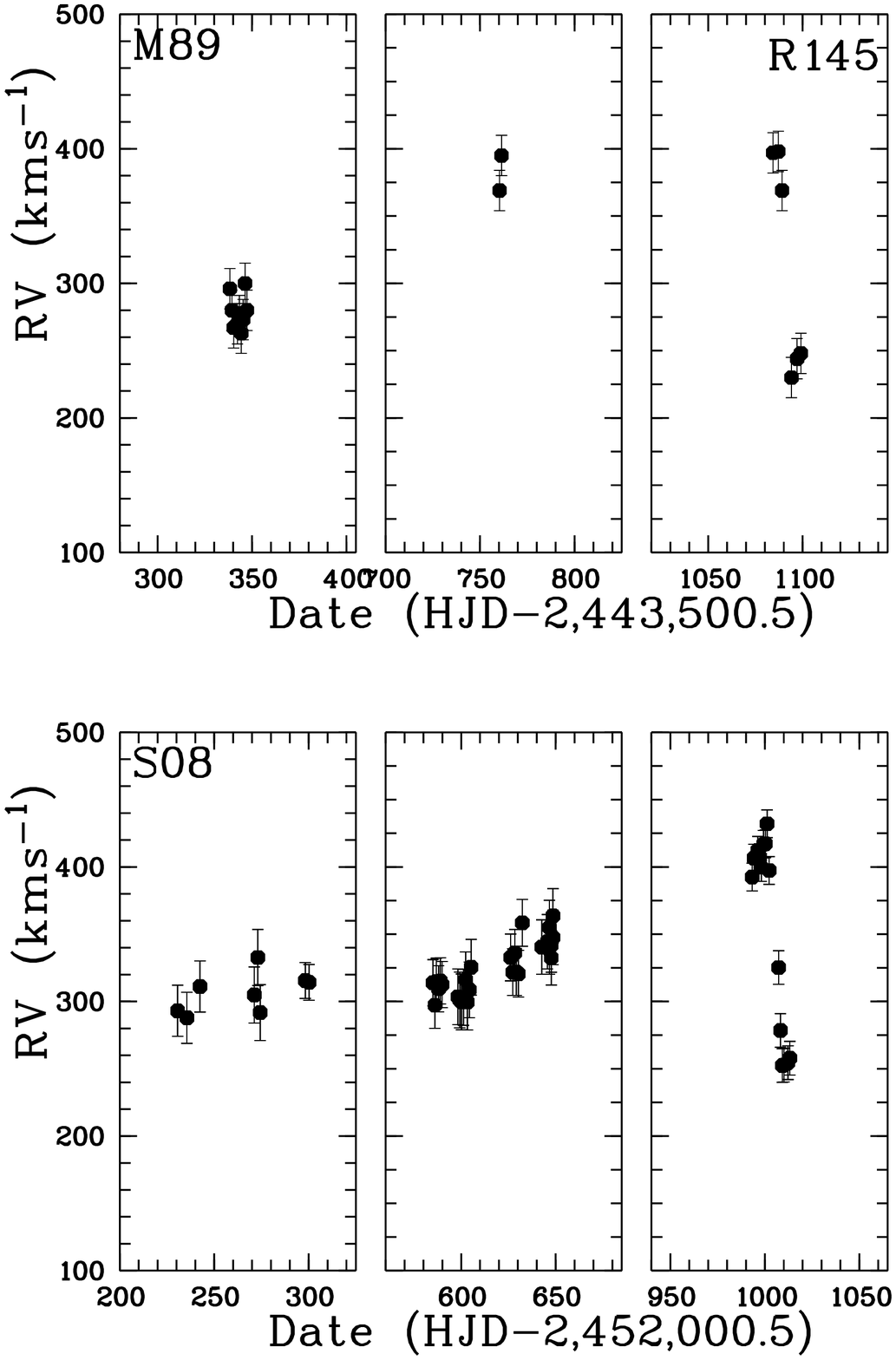}\\
\caption{RV data obtained from He\,\textsc{ii} $\lambda$4686 of R144
(\emph{left}) and R145 (\emph{right}). Shown are the three observing
seasons of M89 (\emph{upper panels}) and S08 (\emph{lower panels}),
respectively.}
\label{rvbatall}
\end{minipage}
\end{figure*}

Fitting a Gauss function to the He\,\textsc{ii} $\lambda$4686 emission
yielded a slightly larger scatter than cross-correlation, but was used
to measure absolute, systemic velocities. These absolute velocities
were then used to correct the relative RVs obtained by
cross-correlation, and to combine our data with those of
M89. Initially, a possible zero-point shift between S08 and M89 was
not corrected for (but see below). Unfortunately, M89's spectra did not
cover He\,\textsc{ii}$\lambda$5412, so that only RV data from
He\,\textsc{ii} $\lambda$4686 and N\,\textsc{iv} $\lambda$4058 could
be combined. The combined RV data for He\,\textsc{ii} $\lambda$4686
are shown in Figure \ref{rvbatall}.

\subsection{Search for Periodicities in the RV Curves}

As reported by S08, a comprehensive period analysis using different
methods of their RV data in the range from 1 to 200 days yielded no
coherent period for R145. We repeated this analysis on the combined
S08+M89 data set, also without success. Additionally, we tried to
apply the phase-dispersion minimization (PDM) method
(\citealt{Stellingwerf78}), which is very reliable because it does not
assume a specific wave form (e.g. sinusoids) and which is therefore
very useful for strongly eccentric binaries with highly distorted RV
curves. Unfortunately however, the data are too dispersed for periods
much longer than 100 days, so the PDM suffers from many severely
depleted or downright empty phase bins. Thus, a different approach was
used.

\subsubsection{R144}

On close inspection in Figure \ref{rvbatall}, R144 does not show any
clearly repeating pattern in S08's RV data, which is the reason why
S08 were not able to establish a period in the first place, and there
certainly is no pattern at all visible in the M89 data set. We
therefore have to conclude that better data are required for R144, if
any periodicity is to be found. Long-term monitoring is being
undertaken.

\subsubsection{R145}

Inspection of R145's RV curves in Figure \ref{rvbatall} reveals a
repeating pattern: In both M89's and S08's last observing season, the
RV curve drops from its highest to its lowest value within a very
short interval of time. This event can consistently be interpreted as
the periastron passage (PP) of a highly eccentric, long-period
binary. But what is its period? We are lucky, because it seems that we
have three such PP events recorded: two in the M89 data set (called
PP1 and PP2), and one in the S08 data set (PP3). We further
distinguish the highest and the lowest RVs during the respective
event; thus, we find that the combined RV data set covers the
following events: PP1 high, PP2 high and low, PP3 high and low. Julian
Dates and RVs for each event are given in Table \ref{PP}.

\begin{table}
\begin{center}
\caption{Measured dates and RVs for the presumed periastron-passage (PP)
events PP1 through PP3 in the data sets of M89 and S08,
respectively. ``High'' and ``low'' refer to maximum and minimum RVs
measured just before and after the supposed PP, respectively. See text for
more details.}
\label{PP}
\begin{tabular}{llll}
\hline
Data set & Event & MJD$^{a}$ & RV$^{b}$\\
         &       &     & [kms$^{-1}$]\\
\hline
M89 & PP1 high  & 44261.4  & 423 $\pm15$\\
\hline
M89 & PP2 high  & 44584.4  & 424 $\pm15$\\
M89 & PP2 low   & 44599.5  & 276 $\pm15$\\
\hline
S08 & PP3 high  & 53001.8  & 432 $\pm15$\\
S08 & PP3 low   & 53012.4  & 252 $\pm15$\\
\hline
$^{a}$JD-2,400,000.5\\
$^{b}$errors are estimated \\
\end{tabular}
\end{center}
\end{table}

It seems that the duration of the PP2 event is somewhat longer than
that of the PP3 event, $\sim$15 versus $\sim$10 days. However, the
observed RV amplitude during PP2 is smaller than that observed during
PP3. It is thus likely that the true minimum RV was missed during the
observations of the PP2 ``low'' event, thus the full RV peak-to-valley
swing of the WR star (i.e., $2K$) might be closer to the value
obtained from PP3, $\Delta RV(PP3) = 180$ kms$^{-1}$.

Measuring the elapsed time between corresponding events yields the
following time intervals, each of which will be an integer multiple of
the period, assumed, of course, there is one:

\begin{eqnarray*}
\Delta T_{\rm 31,hi} = MJD(\rm PP3_{hi}-PP1_{hi}) = (8740.4 \pm 6) days\\
\Delta T_{\rm 32,hi} = MJD(\rm PP3_{hi}-PP2_{hi}) = (8417.4 \pm 6) days\\
\Delta T_{\rm 32,lo} = MJD(\rm PP3_{lo}-PP2_{lo}) = (8412.9 \pm 6) days\\
\Delta T_{\rm 21,hi} = MJD(\rm PP2_{hi}-PP1_{hi}) =  (323.0 \pm 6) days,\\
\end{eqnarray*}

\noindent
where the overall, quadratic error has been obtained from
$\sqrt{4^{2}+4^{2}} \sim 6$. From $\Delta T_{21,hi}$ follows that the
longest possible period is $(323 \pm 6)$ days; otherwise, two high
events would occur during one orbit. The true period is thus an
integer fraction of this period, i.e. $(323 \pm 6) / n$ days, with $n
= 1 .. 5$, since any period shorter than $\sim$60 days is ruled out by
S08's observations; it follows that there are five groups of periods,
$323 \pm 6$, $161.5 \pm 3$, $107.67 \pm 2$, $80.75 \pm 1.5$, and $64.6
\pm 1.2$ days. Within each group, the longer time intervals are used to
refine the results and to obtain more precise periods. Periods which,
within their errors, are consistent with each other, are averaged, and
a quadratic mean error is computed. The resulting periods are listed
in Table \ref{eventtab}.

\begin{table}
\begin{center}
\caption{The five groups of possible orbital periods of R145 as
computed from the condition that the true period has to be an integer
fraction $n$ of the respective time interval $\Delta T_{\rm 21,hi}$,
$\Delta T_{\rm 32,lo}$, $\Delta T_{\rm 32,hi}$, and $\Delta T_{\rm
31,hi}$. For each group, the typical error on the period is
given. Periods that are consistent with each other within these
errors, have been averaged to yield $P_{\rm mean}$ and the mean
error.}
\label{eventtab}
\begin{tabular}{rrrrrr}
\hline
\multicolumn{5}{c}{Possible periods (in days)}  \\
$P_{\rm 21,hi}$ & $P_{\rm 32,lo}$  & $P_{\rm 32,hi}$ & $P_{\rm 31,hi}$ & $P_{\rm mean}$\\[1mm]
\hline
 323.00               &   323.573                & 323.746                &                323.719  &  323.679 \\
   $\pm$6.00          & $\pm$0.231               & $\pm$0.231             &             $\pm$0.222  & $\pm$0.228 \\[2mm]

 161.50               &   161.787                & 161.874                &                161.859  &  161.840 \\
   $\pm$3.00          &   158.734                & 158.819                &                158.916  &  158.823 \\
                      & $\pm$0.115               & $\pm$0.115             &             $\pm$0.111  & $\pm$0.114 \\[2mm]

 107.67               &   106.492                & 106.549                &                106.590  &  106.544 \\
   $\pm$2.00          &   107.858                & 107.915                &                107.906  &  107.906 \\
                      &   109.258                & 109.317                &                109.255  &  109.277 \\
                      & $\pm$0.078               & $\pm$0.078             &             $\pm$0.075  & $\pm$0.077 \\[2mm]

  80.75               &    79.367                &  79.409                &                 79.458  &  79.411 \\
   $\pm$1.50          &    80.123                &  80.166                &                 80.187  &  80.159 \\
                      &    80.893                &  80.937                &                 80.930  &  80.902 \\
                      &    81.679                &  81.722                &                 81.686  &  81.696 \\
                      & $\pm$0.058               & $\pm$0.058             &             $\pm$0.056  &  $\pm$0.057 \\[2mm]

  64.60               &    63.734                &  63.763                &                 63.799  &  63.765 \\ 
   $\pm$1.20          &    64.221                &  64.255                &                 64.268  &  64.248 \\
                      &    64.715                &  64.749                &                 64.744  &  64.736 \\
                      &    65.216                &  65.251                &                 65.227  &  65.231 \\
                      &    65.726                &  65.761                &                 65.717  &  65.735 \\
                      & $\pm$0.047               & $\pm$0.047             &             $\pm$0.043  & $\pm$0.046 \\
\hline
\end{tabular}
\end{center}
\end{table}

\begin{figure}
\includegraphics[width=87mm,angle=0,trim= 5 0 5 0,clip]{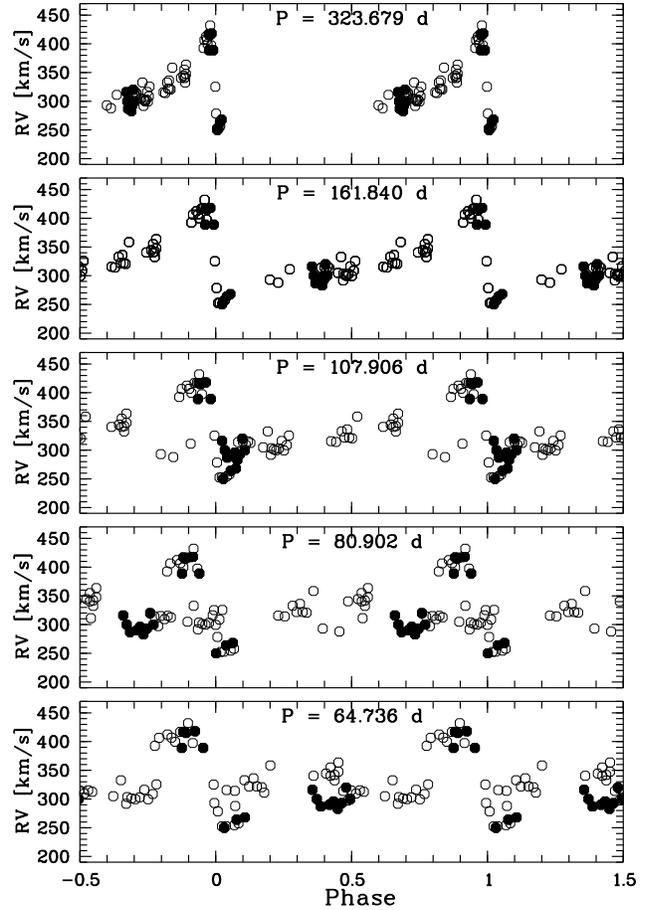}
\caption{Combined RV data of R145 obtained from He\,\textsc{ii}
$\lambda$4686 (M89: filled; S08: empty symbols) and folded into the
corresponding phases of five different, possible periods based on
$P_{\rm mean}$ in Table \ref{eventtab}.}
\label{altperiods}
\end{figure}

Since the data are too sparse, an automated phase-dispersion minimum
method (\citealt{Stellingwerf78}) is not possible; we thus folded the
data into the respective phases and chose the most coherent-looking
curve by eye. This method is of course limited; for instance, there
are no noticeable differences in the RV curves of the 323-day and
161.5-day periods within a given group. Hence, the selected RV curves
shown in Figure \ref{altperiods} are representative. However, the
groups of periods around 107.67, 80.75, and 64.60 days can readily be
ruled out as they yield incoherent RV curves.

The longest possible (and unique) period, $323.679$ days, results in a
very well-defined RV curve and a very eccentric orbit; however, it
does so by construction and, more importantly, shows a data gap of
almost exactly half a phase, which is suspicious. The periods around
$161.5$ days also yield coherent RV curves, but with a much better
phase coverage, and a slightly less eccentric orbit, as can already be
seen by eye.

\subsection{Polarimetric Data}

With these results from the spectroscopy, we proceeded to analyze the
polarimetric data. To derive the orbital parameters, we used the
elliptical orbit model by \citet{Brown82} with the correction by
\citet{SimmonsBoyle84}, modified for an extended source of scatterers
(cf. \citealt{Robert92}), i.e. assuming an ensemble of optically thin,
free electrons that follow the wind density around one of the stars
spherically symmetrically (for details, see \citealt{Moff98}).

For R144, the fit did not converge, and since there is no orbital
solution from the spectroscopy, we were unable to constrain the
fit. This does not entirely come as a surprise, given the small number
of data points ($N=15$) spread over only $\sim$2 weeks, and the rather
small level of variability. Thus, we abandon the study of R144 based
on the current data.

For R145, however, the fit of the polarization data alone yielded a
coherent solution with a period of $P = (159 \pm 1)$ days, which is
consistent with the lower boundary of $P_{\rm 21,hi}$ (see above), but
has consequences that will be discussed below. Encouraged by this
result, we combined the spectroscopic (i.e., M89+S08) and polarimetric
data sets, and forced a simultaneous fit. For cross-verification, we
also used the RV data that were obtained from the slightly weaker, but
less perturbed, He\,\textsc{ii} $\lambda$5411 emission line, with
similar results.

Note that the fits to the respective data sets do not have all orbital
parameters in common. For instance, spectroscopy cannot yield the
inclination angle $i$ and the orientation of the line of nodes
$\Omega$\footnote{We here follow the definition of \citet{Harries96},
i.e. that $\Omega$ is the angle of the ascending node measured from
north through east with the constraint that $\Omega < 180^{\circ}$,
i.e. an ambiguity of 180$^{\circ}$ exists in the determination of
$\Omega$ since it is impossible from polarimetry to discern the
ascending from the descending node. This is done using the
spectroscopic orbit of the binary.}, both of which are calculated from
the polarimetry. Thus, only the following parameters were obtained
from the fit of the combined RV and polarimetric data: Interstellar
polarization $Q_{0}$ and $U_{0}$; $\Omega$ (see above); the orbital
inclination angle $i$; the orbital longitude of the centroid of the
scattering region at periastron passage $\lambda_{p}$ (linked to the
argument of the periastron by the relation $\lambda_{p} = \omega -
90^{\circ}$); the orbital period $P$; the RV amplitude $K_{\rm WR}$ of
the WR component; the systemic velocity $V_{0}$; the systematic RV
shift between the M89 and S08 data sets $\Delta V$; the orbital
eccentricity $e$; the time of periastron passage $T_{0}$; the total
electron scattering optical depth $\tau$ (assuming Thompson scattering
to be the only source of opacity); the exponent $\gamma$ in the power
law describing how the electron density falls off from the WR stars
(e.g., $\gamma = 2$ for an inverse square law around a scattering
point source); the possible systematic shift between the two RV data
sets $\Delta RV = \rm S08-M89$.

\begin{figure*}
\includegraphics[width=115mm,angle=0,trim= 45 0 5 150,clip]{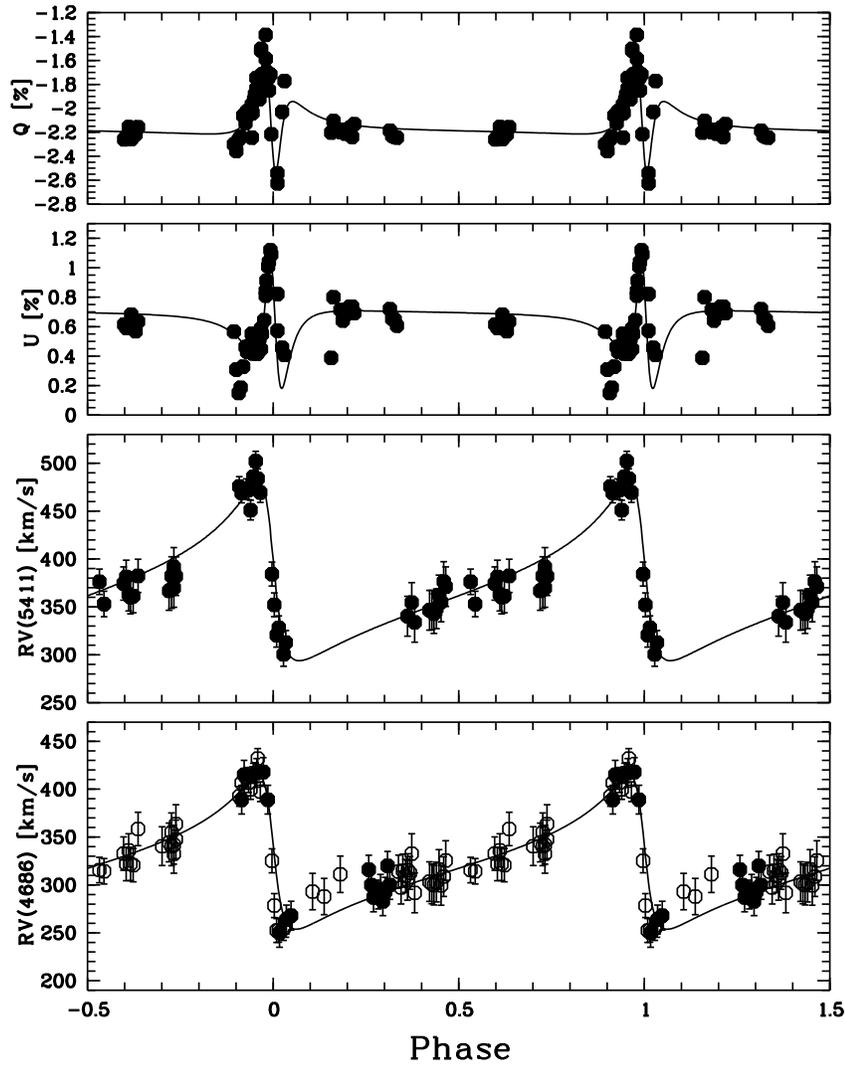}
\caption{Orbital solution for R145 obtained from the simultaneous fit
of the the Stokes $Q$ and $U$, and RV data sets. For the
He\,\textsc{ii} $\lambda$4686 emission line (\emph{bottom panel}), S08
data are shown in filled symbols, and M89 data are shown in empty
symbols. The systematic shift between the two RV data sets, $\Delta
RV$, has been applied. For He\,\textsc{ii} $\lambda$5411, only S08
data could be used for the fit, which is why there are fewer data
points.}
\label{solutions}
\end{figure*}

In order to facilitate convergence, the fit was carried out for three
parameter groups which are, to first order, independent of each other:
1. ($K$, $V_{0}$, $\Delta RV$); 2. ($Q_{0}$, $U_{0}$, $\tau$,
$\gamma$); 3. ($i$, $\Omega$, $\lambda_{p}$, $e$). One group was
fitted while the two others were kept constant. For each permutation
$\chi^{2}$ was computed, and the procedure was repeated until the fit
yielded a constant $\chi^{2}$ for all permutations. In order to
estimate the errors on individual fit parameters, we explored the
sensitivity of the fit (i.e., its $\chi^{2}$) by varying one parameter
around its best-fit value (obtained from the overall fit) while
keeping all other parameters constant. Those variations of each
individual parameter which resulted in a 5\% deterioration of the
$\chi^{2}$ compared to its minimum value from the overall fit, were
used as error levels of the respective parameter. Results of the fits
for both He\,\textsc{ii} $\lambda$4686 and He\,\textsc{ii}
$\lambda$5411 are shown in Table \ref{orbitalelements}. The resulting
orbital solutions together with the polarimetric data folded into the
corresponding phase is shown in Figure \ref{solutions}.

\begin{table}
\begin{center}
\caption{Orbital parameters from the combined fit of both the
spectroscopic and the polarimetric data sets.}
\label{orbitalelements}
\begin{tabular}{l r@{}@{ }p{3.5mm}@{}l r@{}@{ }p{3.5mm}@{}l}
\hline
Parameter &  \multicolumn{3}{c}{He\,\textsc{ii} $\lambda$4686} &  \multicolumn{3}{c}{He\,\textsc{ii} $\lambda$5411} \\
\hline

$\tau$ [\%]                                &     0.048 & $\pm$ &  0.006 &     0.049 & $\pm$ &  0.006 \\ 
$Q_{0}$ [\%]                               &    -2.17  & $\pm$ &  0.04  &    -2.18  & $\pm$ &  0.04  \\
$U_{0}$ [\%]                               &     0.67  & $\pm$ &  0.04  &     0.68  & $\pm$ &  0.04  \\
$\Omega$ [$^{\circ}$]                      &   -40     & $\pm$ &  7     &   -42     & $\pm$ &  7     \\
$i$ [$^{\circ}$]                           &    37     & $\pm$ &  7     &    39     & $\pm$ &  6     \\
$\omega$ = $\lambda_{p} + 90 [^{\circ}]$   &    85.0   & $\pm$ &  6.7   &    83.4   & $\pm$ &  6.6   \\
$P$ [days]                                 &   158.8   & $\pm$ &  0.1   &   158.8   & $\pm$ &  0.1   \\
$K_{\rm WR}$ [kms$^{-1}$]                  &    81     & $\pm$ & 21     &    93     & $\pm$ & 21     \\
$\Delta RV^{a}$ [kms$^{-1}$]               &    10     & $\pm$ & 29     &           &  n/a  &        \\
$e$                                        &     0.70  & $\pm$ &  0.02  &     0.69  & $\pm$ &  0.02  \\
$V_{0}$ [kms$^{-1}$]                       &   328     & $\pm$ & 14     &   382     & $\pm$ & 13     \\
$T_{0}$ [JD-2,450,000.5]                   &  3007.763 & $\pm$ &  0.25  &  3007.763 & $\pm$ &  0.25  \\
$\gamma$                                   &      1.70 & $\pm$ &  0.15  &     1.75  & $\pm$ &  0.14  \\
\hline
$^{a}\Delta RV$=S08-M89
\end{tabular}
\end{center}
\end{table}

The difference in the systemic velocities obtained from
He\,\textsc{ii} $\lambda$4686 and $\lambda$5411 amounts to $\sim$55
kms$^{-1}$. It is well known in the literature that systemic
velocities obtained from emission lines of WR stars display a
systematic red-shift with respect to the true systemic velocity
(e.g. of the host galaxy of the WR star, see e.g. S08). This
phenomenon can be explained by radiative-transfer effects
(cf. \citealt{Hillier89}). However, it is noteworthy that different
lines of the same ionic species will display significantly different
systemic velocities. Cross-verification with R144 yielded the very
same effect with roughly the same shift ($\sim$50 kms$^{-1}$) between
the systemic velocities obtained from the two He\,\textsc{ii} lines,
and shows that the effect is indeed real.


However, in WR+O binaries, He\,\textsc{ii} emission lines are prone to
suffer from profile variations due to excess emission arising in the
wind-wind collision (WWC) region; therefore they might not yield the
correct, but a more or less distorted RV curve (and with this, wrong
orbital parameters). Indeed R145 displays strong excess emissions, and
because R145 is a very eccentric system, the strength of the excess
emission varies considerably over the orbital phase. From simple
considerations, we expect (and find; see Section \ref{WWC} for a more
detailed discussion) the excess emission to be strongest near
periastron. This, however, means that the RVs derived from line
profiles of both He\,\textsc{ii} $\lambda$4686 and $\lambda$5411 are
most affected at an orbital phase where the RV data points have the
most importance for the orbital fit, namely when the RV of the WN
component changes quickly from maximum redshifted to maximum
blueshifted velocities, around periastron passage.

To somewhat reduce the effect of such line-profile variations over the
orbital cycle, we have employed the iterative cross-correlation method
described in S08. In fact, \citet{Foell03a} have shown that using
bisectors to measure the line center yields perfectly similar
results to those obtained by iterative cross-correlation (with
bisectors yielding a slightly larger scatter). Thus, the iterative
form of cross-correlation is relatively robust against the influence
even of strong and variable WWC excess emissions.

\begin{figure}
\includegraphics[width=60mm,angle=-90,trim= 0 10 0 0,clip]{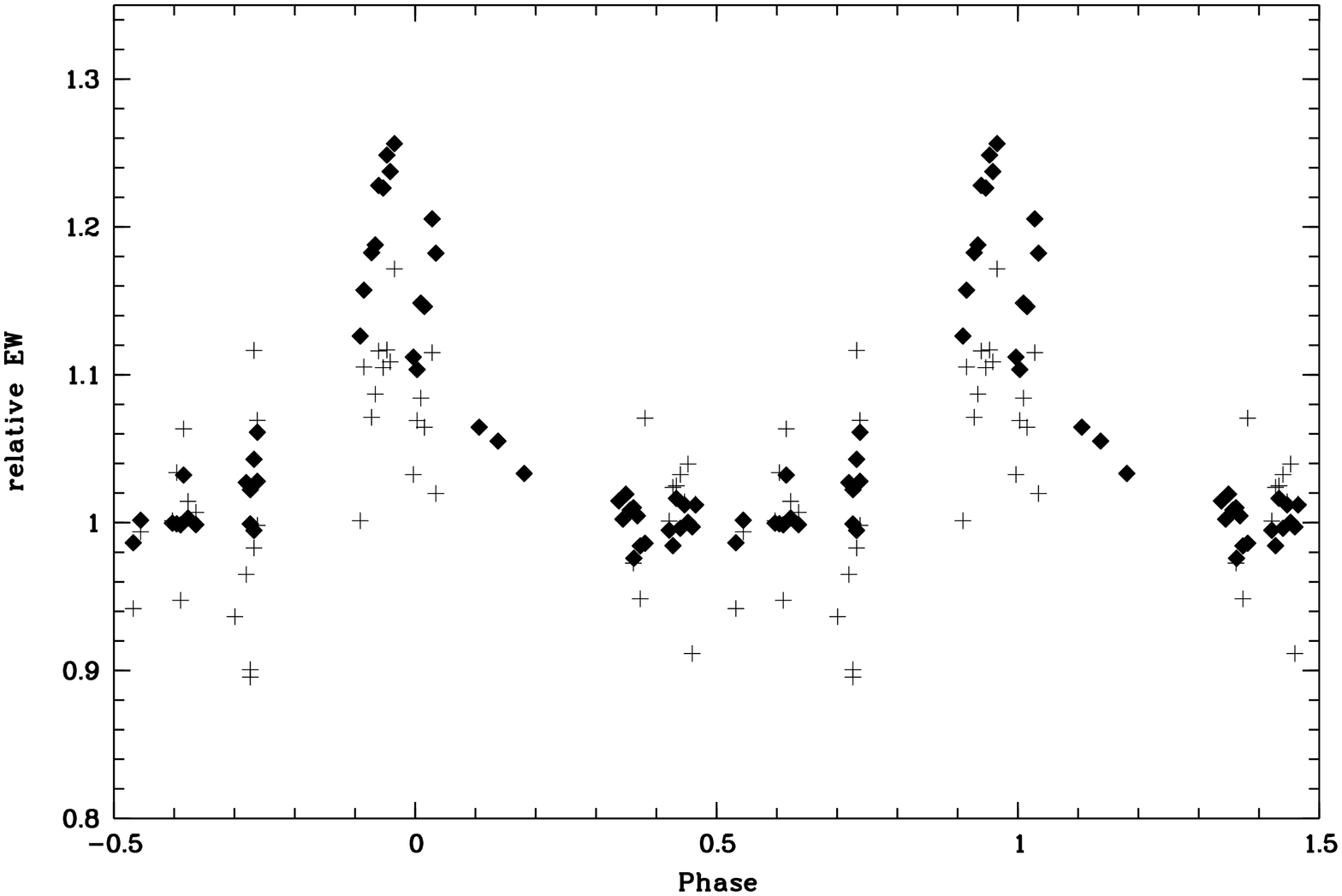}
\caption{Relative EWs (normalized to unity around apastron,
i.e. $\phi=0.5$) of the two emission lines He\,\textsc{ii}
$\lambda$4686 (filled lozenges) and $\lambda$5411 (crosses). While
He\,\textsc{ii} $\lambda$5411 yields more noise because it is the
weaker line, He\,\textsc{ii} $\lambda$4686 suffers from almost twice
the line-strength variability due to increased WWC excess emission
around periastron (i.e. $\phi=0$). Yet both lines yield consistent
orbital parameters.}
\label{compEW}
\end{figure}

Furthermore, we have investigated how the emission-line strength
changes over the orbital cycle due to the variable strength of the
excess emissions. We measured the equivalent widths (EWs) of both
He\,\textsc{ii} $\lambda$4686 and $\lambda$5411, normalized them by
the mean value around apastron (phase interval [0.4;0.6]) when the WWC
excess emissions supposedly are weakest, and folded these relative EWs
into the orbital phase (for $P = 158.8$ days; see Figure
\ref{compEW}). Despite the fact that He\,\textsc{ii} $\lambda$4686
suffers more from the increased WWC excess emission at periastron
(almost twice compared to He\,\textsc{ii} $\lambda$5411), the orbital
solutions the respective lines are fully consistent with each other,
the only difference being the redshifted systemic velocity of
He\,\textsc{ii} $\lambda$5411, as discussed above.

\begin{figure}
\includegraphics[width=60mm,angle=-90,trim= 0 10 0 0,clip]{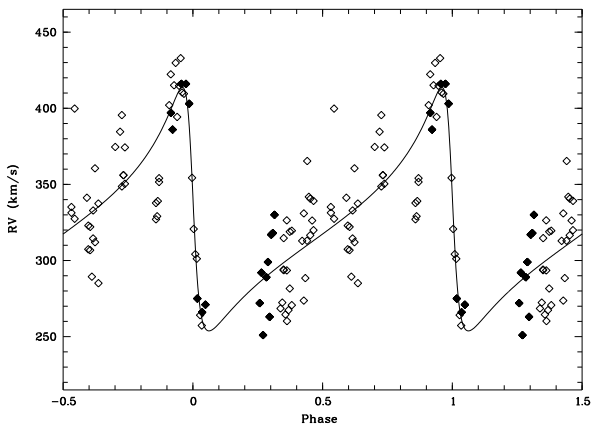}
\caption{Orbital solution as obtained from the He\,\textsc{ii}
$\lambda$4686 emission line (solid line) and RVs obtained from
N\,\textsc{iv} $\lambda$4058 (lozenges; filled symbols are from M89,
empty symbols are based on the S08 data). While the scatter of the RVs
obtained from the N\,\textsc{iv} $\lambda$4058 emission is very large,
there is no indication that the motion of He\,\textsc{ii}
$\lambda$4686 is systematically different.}
\label{compHeII}
\end{figure}

As an additional test, we have used RVs obtained from Gauss-fitting
the weak and narrow, highly-ionized N\,\textsc{iv} $\lambda$4058
emission line. As before, our results were combined with those
published by M89 using the systematic shift between the two data sets
obtained from the orbital fitting of He\,\textsc{ii} $\lambda$4686
($\Delta RV$ of 10 kms$^{-1}$; see Table \ref{orbitalelements}). Due
to its fluorescent origin from ultra-violet transitions
(cf. \citealt{Hillier88}), the N\,\textsc{iv} $\lambda$4058 line is
commonly regarded as very robust against WWC excess emissions, and
usually yields RVs which follow best the true orbital motion of the WR
star. In Figure \ref{compHeII} we have plotted the orbital solution
obtained from He\,\textsc{ii} $\lambda$4686. Overplotted are the
combined RVs obtained from N\,\textsc{iv} $\lambda$4058, shifted by
+110 kms$^{-1}$ to match the mean velocity of the orbital
solution. While the scatter of the data points is expectedly large
(see Section \ref{supertemplate}), which makes it impossible to use
this line for a sensible orbital fit, the points do follow the orbital
motion and, more importantly, do not show a significantly different RV
\emph{amplitude}. Even if this is not fully conclusive evidence, it is
a very plausible indicator that the orbital parameters obtained from
the RVs of the two He\,\textsc{ii} lines are reliable. Moreover, since
the orbital solutions were obtained from a combined set of
spectroscopic and polarimetric data, and the latter is not affected by
line-profile variations, the influence of WWC excess emission is
further attenuated.

Since N\,\textsc{iv} $\lambda$4058 does not suffers from
radiative-transfer effects like the He\,\textsc{ii} lines, and is not
blended, it also yields a relatively precise estimate of the true
systemic velocity of the WR star; \citet{MoffSegg79} found a small,
negative shift for Galactic WN7/WN8 stars, $(-15 \pm 13)$ kms$^{-1}$
(standard deviation). Using the mean velocity of the N\,\textsc{iv}
$\lambda$4058 line as an estimate of the true systemic velocity of the
binary system, we thus obtain $\overline{RV}(4058) = (220 \pm 30)$
kms$^{-1}$. This value is just inconsistent with the reported systemic
velocity of the LMC, $v_{\rm sys} = (280 \pm 20)$ kms$^{-1}$
(\citealt{Kim98}). Either for some reason, N\,\textsc{iv}
$\lambda$4058 has a rather exceptional negative RV, or R145 indeed has
a slightly negative systemic velocity with respect to the LMC.


Based on our error-estimation method, the error for the inclination
angle $i$ was found to be 7$^{\circ}$ and 6$^{\circ}$ for the two
lines He\,\textsc{ii} $\lambda$4686 and $\lambda$5411,
respectively. Choosing a higher cut-off value than the 5\% used will
of course lead to a larger error on $i$ (and all other parameters),
but the $\chi^{2}$ deteriorates very quickly when the inclination
angle varies by much more than 10 to 15$^{\circ}$, and the
discrepancies between the fit and the data become noticeably
large. However, it is well known among polarimetrists that, due to an
asymmetric error distribution, there is a bias of the polarimetric
determination of the inclination angle towards higher values than the
true value of $i$; this bias is more pronounced, the lower the true
inclination angle and the lower the data quality is (in terms of both
the number of available data points and their respective individual
errors); see e.g. Aspin et al. (1981), Simmons et al. (1982) or, more
recently, Wolinski \& Dolan (1994) for a comprehensive discussion of
this issue. Thus, our analysis is bound to \emph{over}estimate the
true value of $i$, which means that the true inclination angle is more
likely lower than $39^{\circ}$, rendering the resulting masses even
higher and less plausible (see below). If, on the other hand, we were
to force a fit with the inclination angle fixed at $90^{\circ}$, the
resulting $\chi^{2}$ would more than double compared to the best-fit
value. Unless the orbital period is wrong altogether, it is thus
unlikely that the inclination angle is that high. This is most
unfortunate, since this result rules out the possibility of R145 being
an eclipsing system, so that photometry does not seem to be a viable
alternative to reveal core eclipses, although atmospheric eclipses may
still be possible. On the other hand, we are not aware of any existing
or published photometry of R145, so that we have no possibility to
verify the polarimetric result. We will therefore adopt the average
value of $i = 38^{\circ} \pm 9^{\circ}$ (quadratic error) for the rest
of the paper, and discuss the consequences for the stellar masses
further below. Given the great importance of an accurate determination
of the inclination angle, we strongly encourage that more high-quality
data be collected.

\subsection{Search for the Companion in R145}

As both S08 and M89 reported, no trace of the companion's absorption
lines could be found in the individual spectra. It would be fairly
obvious if the secondary were an emission-line star as well, like
e.g. in the case of NGC3603-A1 (\citealt{S08a}). This is already an
indication for the secondary being considerably fainter than the
primary, and most likely an absorption-line O star.

In a first attempt to isolate the spectrum of the companion, we
shifted all spectra of R145 into the frame of reference of the WR
component, and co-added them using their respective S/N values as
weights to obtain a high-quality, mean WR spectrum. This mean was then
subtracted from each individual spectrum, thereby removing most of the
WR star's emission lines; thus, if any O-star absorptions were
visible, we expected to find them in the individual residual
spectra. However, there was no obvious trace; any absorption-like
structure was essentially lost in the noise.

We therefore adopted a modified version of the method developed by
\citet{Demers02}, the so-called ``shift-and-add'' method. Originally it
consists in shifting all spectra into the frame of reference of the WR
star and subtracting the mean (which we have done), shifting the
residual spectra back into the frame of reference of the O star, and
co-adding all shifted residuals to obtain a cleaner mean spectrum of
the O star. While this method can be used iteratively to obtain ever
better disentangled spectra of the WR and the O components,
respectively, its caveat is obvious: one has to know the orbit of the
O star, otherwise one cannot shift the residuals into its frame of
reference.

Since we had neither the companion's orbit nor any, even weak,
spectral signatures of absorption lines, we had to assume that $i$)
the companion is indeed an O star and not another emission-line star
(as suggested by the orbital behaviour of the emission-line profiles),
and $ii$) the companion moves in perfect anti-phase (i.e. phase shift
of $\Delta\phi = 180^{\circ}$) to the WR orbit. The orbit of the O
star can then be reconstructed by mirroring the orbital motion of the
WR star. To do so, we considered the \emph{differential} motion
$\delta RV_{\rm ij,WR}$ of the WR star with respect to its systemic
(zero) velocity, as measured at two dates $i,j$, and multiplied it by
$-q$, where $q=K_{\rm O}/K_{\rm WR}=M_{\rm WR}/M_{O}$ the mass ratio
of the binary system, to obtain the corresponding $\delta RV_{\rm
ij,O}$.

For instance, if the WR star changes its RV by 50 kms$^{-1}$, and if,
say, $q=2$, then the O star's RV has to change by $-100$ kms$^{-1}$ in
the same time interval [$i;j$]. i.e. in perfect anti-phase. To reduced
the influence of errors on the individual $RV_{\rm WR}(t)$ measures,
we used the orbital solution of the WR star as obtained above from the
He\,\textsc{ii} $\lambda$4686 line. We emphasize that because this is
an entirely \emph{differential} approach, one does {\bf not} need to
know the systemic velocity of either the WR star or the O star (which
must be different from that of the WR star, not least because most of
the WR emissions display a systematic redshift; see above).

With this tentative $RV_{\rm O}(t)$ curve for the O star, the
individual residual spectra (i.e., after WR subtraction) can now be
shifted into the tentative frame of reference of the O star. Co-adding
of the thus shifted residuals should make the O star's absorption
spectrum appear with greater clarity, \emph{if one has the correct
value for $q$}, because the co-added average spectrum has a higher
S/N. One can then iterate the procedure: The mean O-star spectrum is
subtracted from the original spectra, and the residuals are used to
construct a ``cleaner'' mean WR-star spectrum, i.e. one which now is
relatively free from any O-star absorption lines. With the new mean WR
spectrum, the procedure to obtain the O-star spectrum is repeated, and
so forth.

But how can one know that a chosen $q$ is the right one? If the
residual spectra are shifted by exactly their correct $RV_{\rm O}(t)$
value, the co-added absorption lines will have maximum depths, because
they are co-added while in perfect superposition. If, however, a wrong
mass ratio was assumed (too large or too small), then the co-added
absorption lines will be shallower, because the individual spectra are
not shifted into the correct frame of reference in which they should
stand still over time. Thus, when using different tentative values for
$q$ (or $K_{\rm O}$, which is immediately given by $qK_{\rm WR}$) and
measuring the depth of a given, co-added absorption line, the depth of
the line will go through a maximum when plotted versus $q$. The only
challenge is to wisely choose the O-star absorption lines one
measures, given that the WN star in R145 shows strong emission lines.

We opted for the two lines He\,\textsc{ii} $\lambda$4200 and
$\lambda$4542, because WN emissions at these positions are relatively
weak and because these are the strongest He lines in a hot, early-type
O star, the anticipated companion. We carried out the
``shift-and-add'' exercise scanning through values of $q$ so that the
(intuitively more easily understandable) $K_{\rm O}$ values range from
$-100$ kms$^{-1}$ (i.e., the O star is less massive than the WR star
and moves \emph{in phase} with it, which is of course unphysical, but
it gives a strong constraint on the zero level of the measured
absorption-line depth) to $+600$ kms$^{-1}$ (i.e., the O star is
$\sim8$ times less massive than the WR star and moves in
\emph{anti-phase} with it, just as it should if it was its true
companion), depending on the line. Thus, we also considered the case
were the O star is actually more massive than the WR star, i.e. it
moves with a smaller RV amplitude than the WR star.

\begin{figure}
\includegraphics[width=60mm,angle=-90,trim= 0 10 0 0,clip]{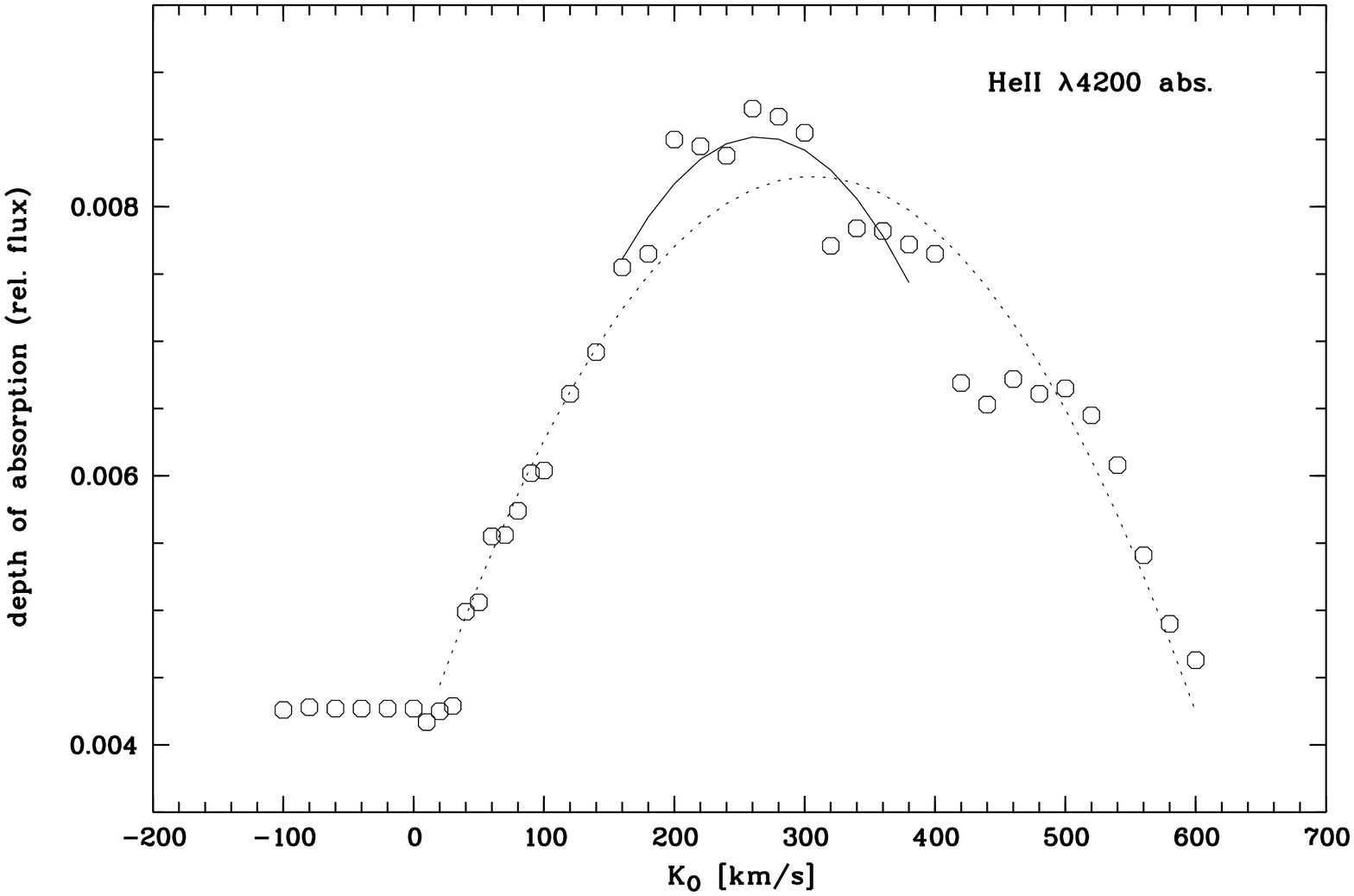}
\includegraphics[width=60mm,angle=-90,trim= 0 10 0 0,clip]{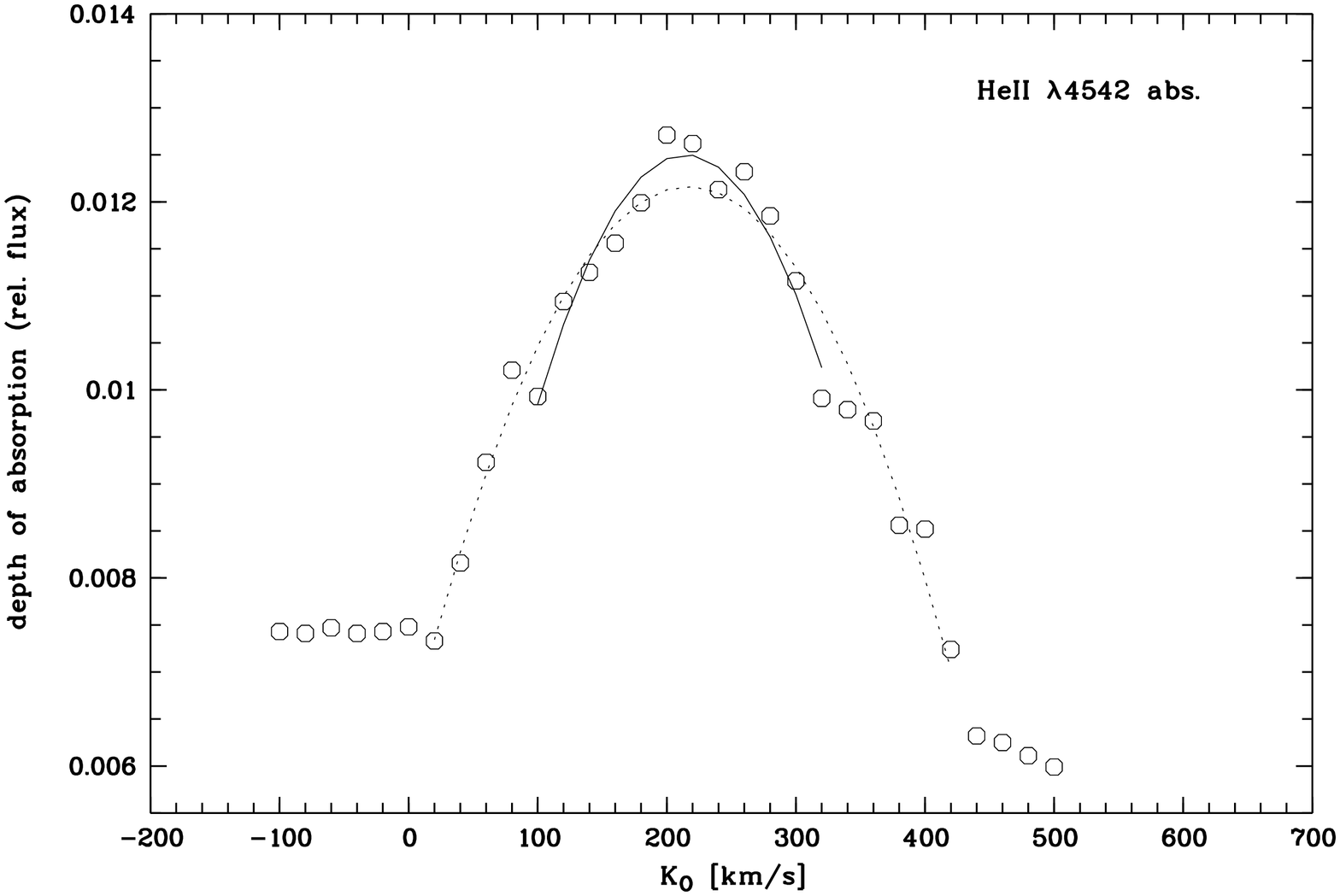}
\caption{Measured depths of the absorption lines in the co-added
spectra versus the assumed value for $K_{\rm O}$, shown for
He\,\textsc{ii} $\lambda$4200 (\emph{upper}) and He\,\textsc{ii}
$\lambda$4542 (\emph{lower panel}). Note the strong asymmetry. The
parabolic fit is shown considering all data (\emph{dotted line}) or
only the central cusp (\emph{solid line}). See text for more details.}
\label{lineshifts}
\end{figure}

From the co-added, mean O-star spectrum, the depths of the
absorption-lines were obtained by hand using the ESO-MIDAS task
GET/GCUR. Continuum values on either side of the absorption were
averaged and subtracted from the peak value such as to obtain positive
values for the line strength, with deeper lines yielding larger
values. Results are shown in Figure \ref{lineshifts} for the
He\,\textsc{ii} $\lambda$4200 and the He\,\textsc{ii} $\lambda$4542
absorptions, respectively, where measured line depths versus the
respective $K_{\rm O}$ value are plotted. The asymmetry of the curve
might be caused by $i$) an intrinsic absorption-line asymmetry due to
a modest P Cygni profile in the O star spectrum, or $ii$) a residual
emission of the WR star due to imperfect subtraction. Both effects
will distort the curve because the intensity is not the same on both
sides of the absorption-line center. Asymmetry is particularly visible
in the data for the He\,\textsc{ii} $\lambda$4200 line; however,
whether this comes from an intrinsic asymmetry in the O-star
absorption or is related to imperfect WR subtraction, cannot be
determined.

\begin{figure}
\includegraphics[width=60mm,angle=-90,trim= 0 0 0 0,clip]{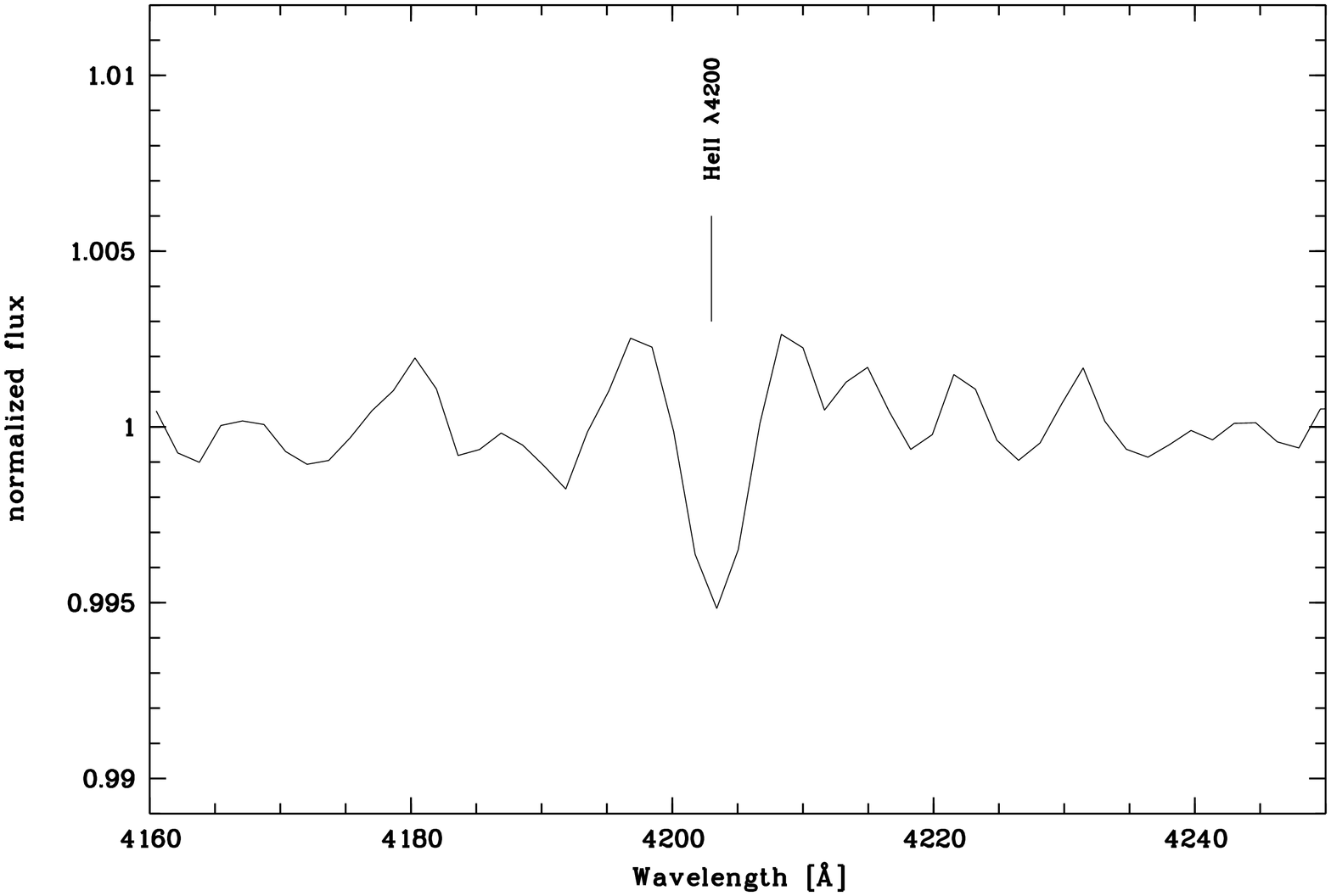}
\includegraphics[width=60mm,angle=-90,trim= 0 0 0 0,clip]{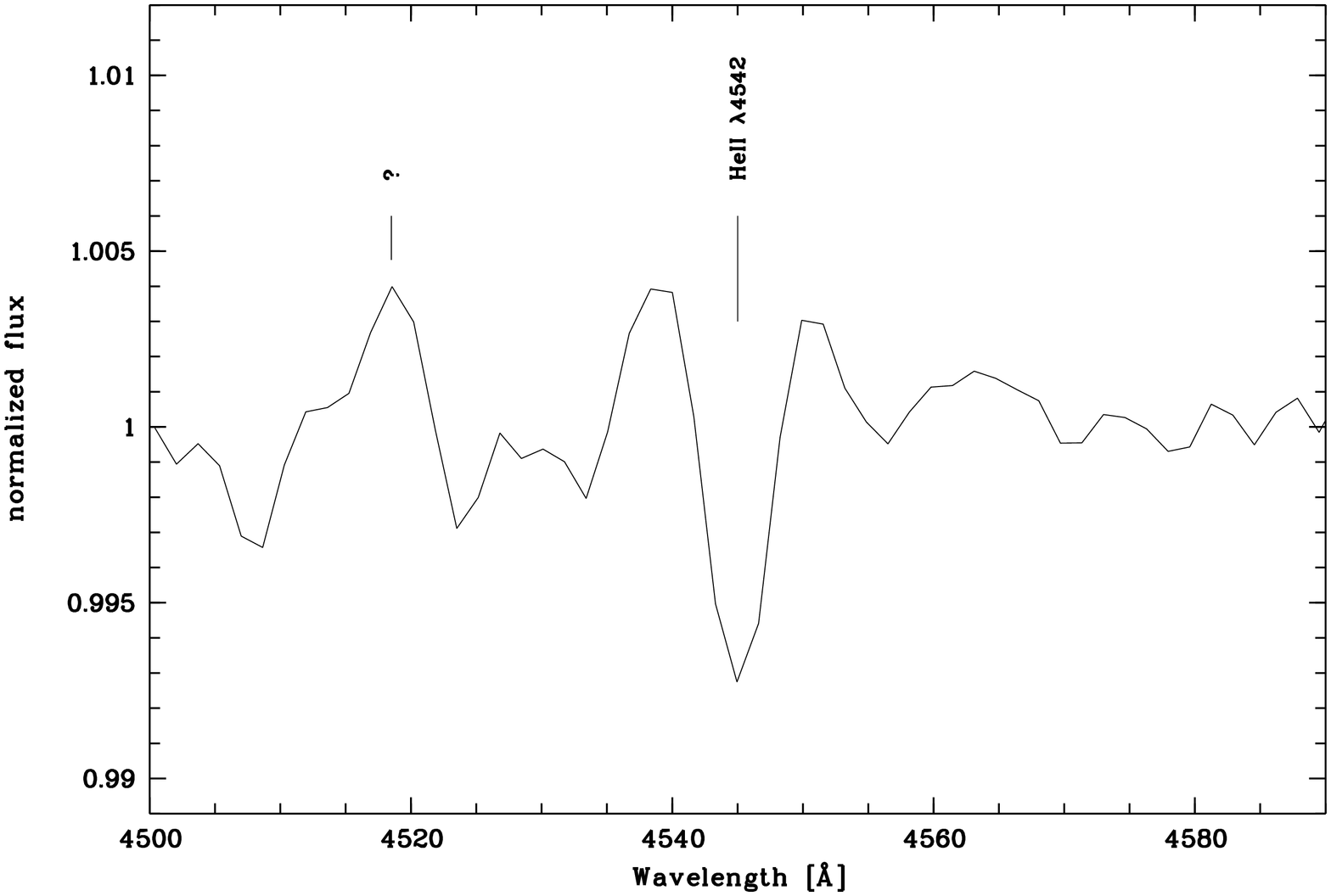}
\caption{Co-added residual spectra for He\,\textsc{ii} $\lambda$4200
(\emph{upper}) and He\,\textsc{ii} $\lambda$4542 (\emph{lower panel})
after being shifted into the O-star's frame of reference, using the
best solution for $K_{\rm O} = 212$ kms$^{-1}$ (see text for more
details). The respective positions of the He\,\textsc{ii} absorption
lines are indicated. Also indicated is an unidentified emission
feature bluewards of He\,\textsc{ii} $\lambda$4542, which seems to be
present in He\,\textsc{ii} $\lambda$4200 as well, but much less
pronounced. Whether this feature is real or merely an artifact of the
shift-and-add method, is unclear.}
\label{ostarlines}
\end{figure}

In order to determine the value for $K_{\rm O}$ at which the
absorption-line depth was maximal, a parabola was fitted to the data
(see Figure \ref{lineshifts}). Comparing He\,\textsc{ii} $\lambda$4200
and He\,\textsc{ii} $\lambda$4542 immediately reveals that the strong
asymmetry in He\,\textsc{ii} $\lambda$4200 leads to a position of the
global maximum which is different from the value obtained from
He\,\textsc{ii} $\lambda$4542, $(306 \pm 69)$ versus $(217 \pm 24)$
kms$^{-1}$. While technically, these two values are consistent with
each other within their respective errors, we re-fitted both sets of
points considering only the central cusp (fit shown in solid line).
We now obtained $(265 \pm 67)$ and $(214 \pm 19)$ kms$^{-1}$ for
He\,\textsc{ii} $\lambda$4200 and He\,\textsc{ii} $\lambda$4542,
respectively. Since the error on the result for He\,\textsc{ii}
$\lambda$4200 is $\sim3.5$ times larger than for He\,\textsc{ii}
$\lambda$4542, this leads to a very small relative weight
($\sim1/12$), if one were to calculate a $\sigma$-weighted
average. Furthermore, S08 reported that due to the lack of
arc-comparison lines towards the blue end of their spectra, RV scatter
was found to be larger than in the central region of the spectra
(i.e., around He\,\textsc{ii} $\lambda$4686). Therefore, we did not
calculate a weighted mean shift but rather entirely relied on the
value obtained from the He\,\textsc{ii} $\lambda$4542 absorption. The
second iteration yielded only a marginal change, $K_{\rm O} = (212 \pm
19)$ kms$^{-1}$. The final, resulting O-star spectrum for the two
absorption lines is shown in Figure \ref{ostarlines}.

Strong artifacts are present, possibly due to imperfect WR
subtraction, not least because the WR emission lines display
phase-dependent excess emission due to wind-wind collisions (see
Section \ref{WWC}). Moreover, upon subtraction of the mean WR
spectrum, slight misalignments between the WR mean and an individual
spectrum will generate pseudo P Cyg profiles in the residuals, whose
strength and width sensitively depend on the amount of misalignment;
at least partially, these artifacts are reduced by the averaging
process. Unfortunately, however, this effectively inhibits one to apply
this method on the He\,\textsc{ii} $\lambda$4686 emission line, which
would be much better visible than any absorption lines, should the
secondary be an emission-line (e.g. an extreme Of/WN6) star as well.

The absorption lines of the putative secondary are very faint; the
depth of the He\,\textsc{ii} $\lambda$4542 absorption line is
$\sim$0.013 in units of relative flux, its ${\rm EW} = (60 \pm 6)$
m\AA~ (error estimated from results obtained for different continuum
positions), and its ${\rm FWHM} = (5.4 \pm 0.2)$ \AA, the latter being
consistent with what is to be expected from a typical O star. Given
the extremely high S/N obtained in the co-added spectra (S/N
$\sim$800), it is thus not implausible that these lines are indeed
real. However, the low spectral resolution of our data, the frequent
rebinning during the shift-and-add, errors in the orbital solution of
the WR star, and noise will degrade the relatively narrow absorption
lines of the O star, so the true depth might be somewhat larger. As an
estimate, we adopt a degradation of the depth by a third, i.e. the
true depth is 1.5 times larger than measured, i.e $\sim$0.02 relative
flux. Comparing with the depth of the absorption lines He\,\textsc{ii}
$\lambda$4200 and $\lambda$4542 in single, early-type O stars, which
are typically 0.12 and 0.16 in continuum units, respectively (e.g.,
\citealt{WalbornFitz90}), we then find that the companion is diluted
by a factor of at least $\sim$8; the WN6h component is thus at least
$\sim$2.2 mag brighter than the O star. The same exercise can be
repeated for the EW. \citet{ContiAlsch71} report a typical
$\log{EW}(4542) = 2.8 \pm 0.1$, i.e. EW(4542) $\sim630^{+170}_{-130}$
m\AA~ for early-type O stars, with little variation between the
spectral types. Using the above measure, we thus obtain a dilution
factor $11^{+2}_{-3}$, consistent with what was derived above. At
$\sim$same bolometric correction, this light ratio directly translates
into a luminosity (and hence, mass) ratio. Using $L \propto
M^{\alpha}$ with $\alpha = 2$ yields the mass ratio of the system $q =
M_{\rm WR}/M_{\rm O} = 3.3^{+0.3}_{-0.5}$.

\subsection{Masses of the Binary Components}

To determine the absolute dynamical masses of both components in R145,
we have averaged the values of the orbital parameters obtained for the
two He\,\textsc{ii} lines $\lambda4686$ and $\lambda5411$ (cf. Table
\ref{orbitalelements}). With the usual relations

\begin{eqnarray*}
M_{1}\sin^{3}i =1.035 \times 10^{-7}
(K_{1}+K_{2})^{2}K_{2}P(1-e^{2})^{3/2} M_{\sun}
\end{eqnarray*}  
and
\begin{eqnarray*}
M_{2}\sin^{3}i =1.035 \times 10^{-7}
(K_{1}+K_{2})^{2}K_{1}P(1-e^{2})^{3/2} M_{\sun},
\end{eqnarray*} 

\noindent
where $P$ is in days, $K$ in kms$^{-1}$, masses $M$ in solar units,
and taking the values $K_{\rm WR} = (87 \pm 21)$ kms$^{-1}$, $e =
0.695 \pm 0.028$, $P = (158.8 \pm 0.1)$ days, and $K_{\rm O} = (212
\pm 19)$ kms$^{-1}$ from the He\,\textsc{ii} $\lambda$4542 absorption
(see above), we obtain minimum masses $M_{\rm WR}\sin^{3}i = (116 \pm
33)$ M$_{\sun}$ for the primary and $M_{\rm O}\sin^{3}i = (48 \pm 20)$
M$_{\sun}$ for the secondary. Thus, the mass ratio is $q = 2.4 \pm
1.2$, consistent with the value that was derived from the
absorption-line strength (see above).

Although the uncertainties are large, minimum masses are already very
high and consistent with R145's absolute visual magnitude. From the
BAT99 catalogue, $v = 12.16$ mag and $b-v = 0.03$. Adopting R144's
$(b-v)_{0} = -0.25$ (\citealt{CroDess98}) for R145, we obtain $A_{v}
\sim 0.87$ mag. With a distance modulus of $DM = 18.5$ mag for the LMC
and the relation $v-M_{v} = DM + A_{v}$, it follows that $M_{v} \sim
-7.2$ mag, i.e. slightly brighter than that of WR20a, $M_{V} = -7.04$
mag (\citealt{Rauw07}),

In WR20a, which consists of two equally massive, and hence equally
bright, components, each one of the components has $M_{V} \sim -6.3$
mag. Since the light from R145 is entirely dominated by the WN6h
component, the brightness difference between the WN6h component in
R145 and one of the WN6ha components in WR20 is $\sim$0.9 mag; this
brightness difference directly translates into a luminosity (and
hence, mass) ratio. The flux ratio between the primary in R145 and one
WN6ha component of WR20a is thus $\sim$2.3; assuming the same
bolometric correction and $L \propto M^{2}$ yields a mass ratio of
$\sim$1.5, i.e. $M_{\rm R145,WN6h} \sim 125 M_{\sun}$. A different
exponent in the mass-luminosity relation is possible, but covered by
the large uncertainties on the stellar mass anyway, as are slight
changes in reddening, intrinsic colors, and distance modulus.

At first glance, this result is in excellent agreement with the
minimum mass obtained from the orbital motion. However, now the
inclination angle has to be taken into account. The fit to the
polarimetric data returns $i = 38^{\circ} \pm 9^{\circ}$, $sin^{3}i =
0.23^{+0.16}_{-0.12}$. Hence, the true masses of the binary components
are higher by a factor $4.3^{+4.5}_{-1.8}$, at least 300 and 125
M$_{\sun}$, respectively. Not only would such high masses be
unprecedented, but R145 would also be severely underluminous for its
mass. If the correct inclination angle were closer to 60$^{\circ}$,
the masses would still be $\sim$1.5 times higher, but then the
resulting absolute masses would be roughly (if just) consistent with
the intrinsic brightness of the system. However, the $\chi^{2}$ of the
fit quickly deteriorates if the inclination angle is 10 to
15$^{\circ}$ larger than the optimum value (see above), so there seems
to be no margin for a significantly higher inclination angle.

Unless we accept that the polarimetry yields an entirely wrong
inclination angle (which then would question the validity of the
orbital solution altogether), the easiest way out is that the
shift-and-add method is at fault, and that the secondary has a smaller
RV amplitude, i.e. the mass (and hence, luminosity) ratio is smaller
(A \emph{larger} RV amplitude of the secondary would not remedy the
situation, because the resulting system mass would be even higher, if
everything else remained the same.) Then, however, the question arises
why the secondary is not visible in the composite spectrum, let alone
during the shift-and-add process. An explanation might be that the
secondary itself is massive (luminous) enough to display a significant
stellar wind, and resembles e.g. an extreme Of/WN6 star. The
He\,\textsc{ii} $\lambda$4686 emission of such a star would be weak
enough to be dominated by the much stronger line of the WN6h
component, and possibly even be drowned out by excess emission arising
from wind-wind collision in the residual spectra (see Section
\ref{WWC}); on the other hand, the He\,\textsc{ii} $\lambda$4200 and
$\lambda$4542 absorptions could (at least partially) be filled up by
wind emission. The latter would seriously affect the usefulness of
measuring the depth or the EW of the absorption lines in the co-added
residuals (any further challenges with this notwithstanding, see
above), and entirely falsify the luminosity (and hence, mass) ratio
that was derived from the absorption-line strengths, etc. In the
following section, we will therefore use an independent approach to
obtain an estimate on the inclination angle.

\subsection{Wind-Wind Interaction Effects}
\label{WWC}

As mentioned above, it is well known that in WR+O binaries, where both
stars have strong winds, WWCs occur which generate two cone-shaped
shocks separated by a contact discontinuity. This contact
discontinuity wraps around the star with the weaker wind (i.e. that
with less wind momentum), which is usually the O star. As the shocked
matter flows with high velocities downstream along the contact
discontinuity, it cools, giving rise to excess emissions which can be
seen atop some optical emission lines of the WR star. The position of
the excess bumps atop the emission lines depends on the orientation of
the flow. The shock-cone acts like a beacon of a lighthouse: if it is
pointing towards the observer, the excess emission will be
blue-shifted, and half an orbit later, the matter will be flowing away
from the observer, resulting in a red-shifted excess emission. Due to
the orbital motion, the Coriolis force will slightly distort the cone,
tilting its axis and making it curve away at large separations, with
respect to the line of sight between the WR and the O star.

\begin{figure}
\includegraphics[width=85mm,angle=-0,trim= 0 0 0 0,clip]{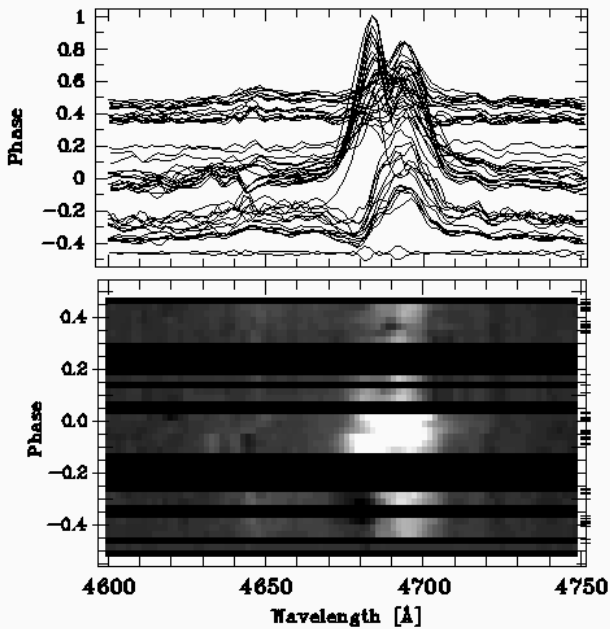}
\caption{Overplot ({\it upper panel\/}) and greyscale ({\it lower
panel\/}) of the residual spectra in the observer's frame of
reference, after subtraction of the minimum profile. While the
He\,\textsc{ii} $\lambda$4686 emission line is highly variable, the
strongest increase occurs around periastron passage ($\phi=0$). The WR
star passes in front at $\phi = 0.5$ and it can (if just) be seen that
it is at this phase that the WWC excess emission displays the largest
redshift. Horizontal black stripes in the greyscale correspond to
empty phase bins.}
\label{mon119}
\end{figure}

To study this behavior in more detail in R145, all spectra were
shifted into the rest frame of the WR star. Then, a minimum spectrum
was computed by averaging spectra obtained near apastron, i.e. around
$\phi\sim0.5$, when the WWC excess emissions are weakest. (While the
excess emissions are not zero, as will be shown below, differences
remain small and have no influence on the conclusions drawn here.) Due
to the eccentricity of the orbit, phases near apastron are well
covered, and a good-quality minimum spectrum was obtained. The minimum
was subtracted from all spectra of the time series, the residuals were
shifted back into the observer's frame of reference, folded into the
corresponding phase of the binary period, and a dynamic spectrum was
constructed where the intensities are coded in greyscales, white being
the strongest emission. Forty phase-bins were used for good
resolution, covering one orbital cycle centered on $\phi=0$, at which
periastron passage occurs. The result is shown in Figure \ref{mon119}.

Since the WWC excess emissions vary in shape and position over the
orbital cycle in a defined way, modelling their profile can be used to
obtain an independent measure of the system parameters like e.g. the
orbital inclination (see \citealt{Luehrs97}). Unfortunately, the
quality of our data is not sufficient to grant such an analysis, but a
simplified method was developed by \citet{Hill00} that still allows one
to obtain valuable information from the phase dependency of the excess
emission.

\begin{figure}
\includegraphics[width=90mm,angle=0,trim= 0 0 0 0,clip]{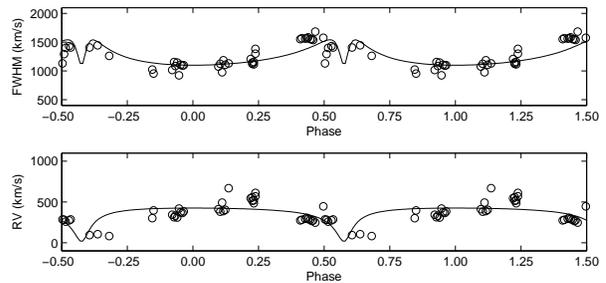}
\caption{Fit to the FWHM (\emph{upper panel}) and mean RV (\emph{lower
panel}) of the WWC excess-emission profile. See text for more details.}
\label{WWCfit}
\end{figure}

Following Hill et al.'s (2000) approach, we have measured the
full-width at half maximum (FWHM) and the mean RV of the
excess-emission profiles, i.e. of the residuals after having
subtracted the minimum spectrum (see above). These values follow the
relation

\begin{eqnarray*}
{\rm FWHM} &=& C_{1} + 2v_{\rm cone}\sin\theta\sqrt{1-\sin^{2}i\cos^2(\phi-\delta \phi)}\\
{\rm RV} &=& C_{2} + v_{\rm cone}\cos\theta \sin i\cos(\phi-\delta \phi),\\
\end{eqnarray*}

\noindent
with $C_{1}$ and $C_{2}$ simple constants, $v_{\rm cone}$ the stream
velocity along the shock cone, $\theta$ the half opening angle of the
cone, $\phi$ the respective orbital phase\footnote{$\phi$ is for a
circular orbit; for an elliptical orbit, we replace $\phi$ by $v +
\omega - 90^{\circ}$, where $v$ is the true anomaly. Thus, the WR star
passes inferior conjunction when $v + \omega = 90^{\circ}$.}, $\delta
\phi$ the tilt angle of the cone axis due to the Coriolis force which
leads to a phase shift between the orbital and the cone motion, and
$i$ the orbital inclination angle (also see Fig. 6 of
\citealt{Hill00}). A constraint is that the stream velocity in the
cone must not exceed the wind terminal velocity (1300 kms$^{-1}$, see
Section \ref{massloss}). A least-square fit to the data is shown in
Figure \ref{WWCfit}. It is satisfactory for the FWHM, but less so for
the mean RV, most likely to insufficient data quality. We obtain
$C_{1} = 30$ kms$^{-1}$, $C_{2} = 220$ kms$^{-1}$, $v_{\rm cone} =
810$ kms$^{-1}$, $\theta = 69^{\circ}$, and $\delta \phi = 0$. The
inclination angle is found to be $i = 45^{\circ}$, in excellent
agreement with the result obtained by polarimetry. Just as in the case
of the polarimetry (see above), a fit with $i=90^{\circ}$ is ruled out
by the data; thus it appears that R145 is indeed seen under such a low
inclination, with the consequences discussed above.

As can be seen from Figure \ref{mon119}, the WWC excess emission is by
far strongest at periastron passage, i.e. when the two stars are
closest to each other, and quickly drops when the separation between
the stars increases. The dependence of the excess emission on the
separation can be studied in more detail, using the recipe by
\citet{Stevens92}. These authors have shown that the WWC zone of
relatively wide binaries can be treated as adiabatic, even at
periastron; therefore the amount of X-ray flux generated by WWC is
expected to go with the inverse separation between the stars (also see
\citealt{Usov92}). Since He\,\textsc{ii} $\lambda$4686 is a
recombination line (i.e. formed through ``collision'' of two
particles) the formalism of \citet{Stevens92} can be applied to the
strength of the excess emission, too. Assuming the WWC excess emission
in He\,\textsc{ii} $\lambda$4686 to be optically thin, we can thus
expect a rise and fall of the emissitivity which is $\propto D^{-1}$,
where $D$ is the separation between the WR and the O star. We here use
the relative separation $D(v)$ as a function of true anomaly $v$,
normalized by the minimum separation $d_{\rm min}$, i.e.

\begin{eqnarray*}
D(v) = d(v)/d_{\rm min} = \frac{1+e}{1+e\cos v},
\end{eqnarray*}

\noindent
where $v$ is the true anomaly of the orbital ellipse and $D_{\rm min}
= 1$ at periastron ($v=0$, phase of the elliptical orbit $\phi=0$),
and consequently $D_{\rm max} = (1+e)/(1-e)$ at apastron ($v=\pi$,
$\phi=0.5$). Note that due to this normalization, the expression is
independent of the inclination angle. However, while the excess
emission is minimum at apastron, it is not generally zero. Thus, the
observed EW of the He\,\textsc{ii} $\lambda$4686 emission is at any
time the sum of the EW of the unperturbed WR emission, $E_{0}$, and a
separation-dependent excess flux, $X_{0}/D(v)$, so that $EW(D) = E_{0}
+ X_{0}/D(v)$, with $X_{0}$ a constant (the maximum excess), and
$D(v)$ as defined above. For verification, we also fitted a steeper
dependency, $EW'(D) = E'_{0} + X'_{0}/D(v)^{2}$.

\begin{figure}
\includegraphics[width=60mm,angle=-90,trim= 0 0 0 0,clip]{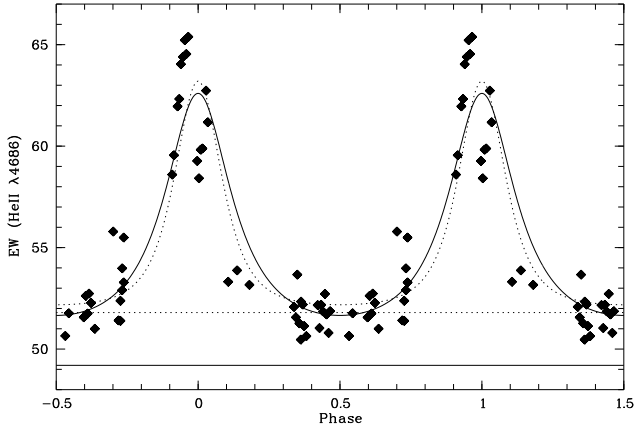}
\caption{Equivalent widths of the He\,\textsc{ii} $\lambda$4686
emission as a function of orbital phase (i.e. separation
$D$). Overplotted are the hypothetical emissivity curves; the solid
line represents an emissivity rising and falling with $D^{-1}$,
whereas the dotted line is for a steeper $D^{-2}$ law. The horizontal
lines indicate the unperturbed equivalent width, as obtained from the
fit for the $D^{-1}$ case (solid line) and the $D^{-2}$ case (dotted
line). See text for more details.}
\label{ew4686}
\end{figure}

From equal-weight least-square fitting, we obtained pairs of
$(E_{0};X_{0}) = (49.2;13.4)$ for the $1/D$ dependency, and
$(E'_{0};X'_{0}) = (51.8;11.4)$ for the $1/D^{2}$ case (EWs in
\AA). The data are shown in Figure \ref{ew4686}, with the two
theoretical curves overplotted. While both curves certainly look
equally well-fitting, the $1/D^{2}$ curve yields a $\chi^{2}$ that is
$\sim$10\% larger than that for the $1/D$ case. Since our data are
rather noisy, and show larger scatter around periastron than around
apastron, it is difficult to assess the true (\emph{a posteriori})
measurement error per individual data point. Using the scatter of the
data around apastron for this, the normalized $\chi{2}$ of the fit is
$\sim$2, but larger if the scatter around periastron is adopted.


\subsection{The Mass-Loss Rate of the WN Star}
\label{massloss}

Independent of any issues regarding the orbital inclination angle, we
can obtain a rough estimate of the mass-loss rate, $\dot{M}_{\rm WR}$,
of the WN6h star in R145, assuming that the companion has a negligible
mass loss. Following the prescription of Moffat et al. (1998; also see
references therein), we use the total optical depth in polarization
due to electron scattering

\begin{eqnarray*}
\tau = 23\sigma_{T}f_{c}\dot{M}_{\rm
WR}(1+\cos^{2}i)\sin i/[(16\pi)^{2}m_{p}v_{\infty}a],
\end{eqnarray*}

\noindent
where $\sigma_{T}$ is the Thomson electron-scattering cross-section,
$f_{c}=I_{\rm O}/(I_{\rm WR}+I_{\rm O})$ the intensity ratio between
the O star and the total luminosity, $m_{p}$ the proton mass,
$v_{\infty}$ the wind terminal velocity, and $a = [P^{2}(M_{\rm
WR}+M_{\rm O})]^{1/3}$ the orbital separation. For $f_{c}$, we use the
light ratio $\sim0.125$ as derived above from the absorption-line
strength of the O star companion with respect to single early-type O
stars. From tailored atmosphere analysis of R144, \citet{CroDess98}
found $v_{\infty}=1350$ kms$^{-1}$. S08 report that the FWHM of the
He\,\textsc{ii} $\lambda$4686 emission line in R145 is only slightly
smaller than in R144; assuming that both stars are similar enough that
the FWHM of He\,\textsc{ii} $\lambda$4686 can act as a reasonable
proxy of the terminal wind velocity, we adopt for R145 a value of
$v_{\infty}=1300$ kms$^{-1}$. For $i$ and $\tau$, the averaged values
from Table \ref{orbitalelements} are taken.

Inserting those values in the equation given above yields a very high
mass-loss rate of $\dot{M}_{\rm WR} = 1.0 \times 10^{-4}$
M$_{\sun}$yr$^{-1}$ (remarkably, for inclination angles between
$40^{\circ}$ and $90^{\circ}$ this value changes barely). From
tailored analysis and using models without iron-line blanketing and
wind clumping, \citet{CroDess98} found $\dot{M} = 1.5 \times 10^{-4}$
M$_{\sun}$yr$^{-1}$ for R144. Introducing typical clumping factors $f
\sim 0.1$ usually reduces the mass-loss rates derived from
spectroscopic means by a factor of $1/\sqrt{f} \sim 3$, i.e. to more
sensible $\dot{M} = 5 \times 10^{-5}$ M$_{\sun}$yr$^{-1}$; since R144
is the more luminous star of the two, one would expect an even lower
mass-loss rate for R145. However, since polarization depends only on
the number of scatterers (i.e. electrons), polarimetric diagnostics
are unaffected by wind clumping, and polarization is indeed observed
to be that high in R145.

Thus, assuming the models to be correct, such a high, continuous
mass-loss rate for R145 would be sensible only if the wind were
particularly smooth. Another way to lower the derived mass-loss rate
would be to considerably decrease the orbital separation $a$ or the
terminal wind speed $v_{\infty}$. The former would require a
mass-ratio closer to unity, with the consequences discussed above,
while the latter is unlikely, given the close resemblance of R145 to
R144. We thus conclude that the primary of R145 may in fact have a
very high mass-loss rate, subject to further confirmation.


\section{Summary and Conclusion}
\label{section4}

We have combined previously published radial velocity (RV) data from
\citet{M89} with data obtained by \citet{S08b}, along with previously
unpublished polarimetric data, for the two WN6h stars R144 and R145 in
the LMC. While the former star was first suspected to be binary by
\citet{S08b}, the latter had already been identified as a binary by
Moffat (1989).

While our study could not reveal any periodicity in the data of R144,
we have, for the first time, established the full set of orbital
parameters for R145. The orbital period was found to be $P = (158.8
\pm 0.1)$ days, in contrast to the preliminary estimate of 25.4 days
by Moffat (1989), based on the assumption of a circular orbit. From
our analysis of the polarimetric data for R145, the inclination angle
of the orbital system was found to be very low, $i = 38^{\circ} \pm
9^{\circ}$. This value was confirmed by a simplified study of the
phase-dependent excess emission on the He\,\textsc{ii} $\lambda$4686
line due to wind-wind collision. Following the approach of
\citet{Hill00}, we obtained a consistent value of $i = 45^{\circ}$.

By applying a modified version of the shift-and-add method originally
developed by Demers et al. (2002), we were able to isolate the
spectral signature of the O-star companion. We found the RV amplitude
of the primary (WN) star to be $K_{\rm WN} = (87 \pm 21)$ kms$^{-1}$,
while the RV amplitude of the secondary (O) star was found to be
$K_{\rm O} = (212 \pm 19)$ kms$^{-1}$. We derived the \emph{minimum}
masses of the WR and the O component to be, respectively, $M_{\rm
WN}\sin^{3}i = (116 \pm 33)$ M$_{\sun}$ and $M_{\rm O}\sin^{3}i = (48
\pm 20)$ M$_{\sun}$. While the large mass ratio $q = M_{\rm WR}/M_{\rm
O} \sim 2.4 \pm 1.2$ is compatible with the strong dilution of the
O-companion's absorption lines, the absolute masses of the two stars
are problematic. If the derived, low inclination angle is correct, one
obtains minimum absolute masses of 300 and 125 M$_{\sun}$ for the WN
and the O star, respectively, but such high masses are not compatible
with the observed luminosity of R145, which puts the system between
NGC3603-A1 (see \citealt{S08a}) and WR20a (\citealt{Rauw07}). Given that
both NGC3603-A1 and WR20a consist of two $\sim$equally bright stars,
while in R145 the WN component dominates the light, masses of 120 and
50 M$_{\sun}$ are more likely, but this would require an inclination
angle close to 90$^{\circ}$, which is not supported by the
data. Another possibility, the mass ratio being closer to unity, is not
supported by the results of the shift-and-add method. Thus, something
is clearly odd in R145, and more and better data are required.

From the polarimetric data for R145, we also estimated a rather
extreme value of the mass-loss rate for the WR component, of $\sim
10^{-4}$ M$_{\sun}$yr$^{-1}$, which is suspiciously high even for such
a massive and hence luminous object. Analyzing the He\,\textsc{ii}
$\lambda$4686 excess emission from wind-wind collisions in R145, we
find that its strength depends on the separation $d$ of the two stars
as $d^{-1}$, as is expected from theory of adiabatic winds, but that a
steeper dependence, $\propto d^{-2}$, which was reported for WR140 by
\citet{Marchenko03}, cannot unambiguously be ruled out.

Given the very large uncertainties on the values derived from our
data, our findings cannot be considered other than
preliminary. However, given the fact that R145 seems to contain one of
the most luminous and thus probably also most massive stars known in
the Local Group, we feel that any additional effort is well deserved
to obtain more reliable results. Therefore, we are currently carrying
out a long-term monitoring of this system, along with R144, but we
strongly encourage independent studies of these potential cornerstone
objects.

\section*{Acknowledgments}

OS would like to thank Paul Crowther for fruitful discussions and the
referee, Otmar Stahl, for comments that helped to improve this
paper. AFJM and NSL are grateful for financial aid to NSERC
(Canada) and FQRNT (Qu\'ebec).


\label{lastpage}

\end{document}